\documentclass[trackchanges,default]{aastex701}

\usepackage{amsmath}
\usepackage{amssymb}
\usepackage{bm}
\usepackage{CJK}
\usepackage{comment}
\usepackage{graphicx}
\usepackage{natbib}
\usepackage{multirow}
\usepackage{upgreek}
\usepackage{hyperref}

\newcommand{\matnot}[1]{\mathbf{#1}}

\submitjournal{The Open Journal of Astrophysics via arXiv.}

\begin{document}

\title{A Model to Map Brown Dwarf Companions to White Dwarfs in Partial Secondary Eclipse}

\author[orcid=0000-0002-7139-3695, gname={Arthur}, sname={Adams}]{Arthur D. Adams}
\affiliation{Department of Astronomy, University of Virginia, 530 McCormick Road, Charlottesville, VA 22904, USA}
\email[show]{gba8kj@virginia.edu}

\author[orcid=0000-0003-2969-6040, gname={Yifan}, sname={Zhou}]{Yifan Zhou}
\affiliation{Department of Astronomy, University of Virginia, 530 McCormick Road, Charlottesville, VA 22904, USA}
\email{uyr2ce@virginia.edu}

\author[orcid=0000-0003-3714-5855, gname={Daniel}, sname={Apai}]{Daniel Apai}
\affiliation{Steward Observatory, University of Arizona, 933 N. Cherry Ave., Tucson, AZ 85721, USA}
\affiliation{Lunar and Planetary Laboratory, University of Arizona, 933 N. Cherry Ave., Tucson, AZ 85721, USA}
\affiliation{James C. Wyant College of Optical Sciences, University of Arizona, Tucson, AZ 85721, USA}
\affiliation{Earth, Atmospheric, and Planetary Sciences, Massachusetts Institute of Technology, 77 Massachusetts Avenue, Cambridge, MA 02139, USA}
\email{apai@arizona.edu}

\author[orcid=0009-0005-9531-7733, gname={Daphne}, sname={Broski-Laing}]{Daphne Broski-Laing}
\affiliation{Department of Astronomy, University of Virginia, 530 McCormick Road, Charlottesville, VA 22904, USA}
\email{rqy2qb@virginia.edu}

\author[orcid=0000-0003-2478-0120, gname={Sarah}, sname={Casewell}]{Sarah L. Casewell}
\affiliation{Centre for Exoplanet Research, School of Physics and Astronomy, University of Leicester, University Road, Leicester LE1 7RH, UK}
\email{slc25@leicester.ac.uk}

\author[orcid=0000-0003-0825-4876, gname={Jenni}, sname={French}]{Jenni R. French}
\affiliation{School of Physics and Astronomy, University of Birmingham, Edgbaston, Birmingham B15 2TT, UK}
\email{j.french.4@bham.ac.uk}

\author[orcid=0000-0002-4321-4581, gname={Laura}, sname={Mayorga}]{L. C. Mayorga}
\affiliation{Johns Hopkins University Applied Physics Laboratory, 11100 Johns Hopkins Road, Laurel, MD 20723, USA}
\email{laura.mayorga@jhuapl.edu}

\author[orcid=0009-0003-6759-7067, gname={Lael}, sname={Shin}]{Lael Shin}
\affiliation{Steward Observatory and Department of Astronomy, The University of Arizona, 933 N. Cherry Avenue, Tucson, AZ 85721, USA}
\email{lshin@arizona.edu}

\author[orcid=0000-0003-2278-6932, gname={Xianyu}, sname={Tan}]{Xianyu Tan}
\affiliation{Tsung-Dao Lee Institute \& School of Physics and Astronomy, Shanghai Jiao Tong University, 1 Lisuo Road, Shanghai 201210, China}
\email{xianyut@sjtu.edu.cn}

\author[orcid=0000-0003-1487-6452, gname={Ben}, sname={Lew}]{Ben W. P. Lew}
\affiliation{Bay Area Environmental Research Institute, Moffett Field, CA 94035, USA}
\affiliation{NASA Ames Research Center, Moffett Field, CA 94035, USA}
\email{weipeng1992@hotmail.com}

\author[orcid=0000-0003-3667-8633, gname={Joshua}, sname={Lothringer}]{Joshua D. Lothringer}
\affiliation{Space Telescope Science Institute, 3700 San Martin Drive, Baltimore, MD 21218, USA}
\email{jlothringer@stsci.edu}

\author[orcid=0000-0001-9521-6258, gname={Vivien}, sname={Parmentier}]{Vivien Parmentier}
\affiliation{Laboratoire Lagrange, Observatoire de la C\^{o}te d'Azur, Universit\'{e} C\^{o}te d'Azur, Nice, France}
\email{vivien.parmentier@oca.eu}

\begin{abstract}
Eclipse mapping is a well-established technique for constraining the spatial emission of hot Jupiter exoplanets. Brown dwarfs in ultra-short orbits around white dwarfs are an analog to hot Jupiters that can only ever be eclipsed within a limited range of latitudes. We adapt the eigenmapping method to disentangle the signals from the ever- and never-occulted regions. The result of this signal separation is a super-position of map structures, yielding sharp boundaries at the occultation region. We introduce a model that penalizes the harmonic complexity of the model map with Tikhonov regularization, whose parameters are tuned to reproduce the fit quality of the eigenmapping model. We use JWST NIRSpec/PRISM full-orbit observations of the eclipsing WD 1202-024 system as a template to construct and run injection-recovery tests on simulated light curves, isolating both latitudinal structures as well as emission from a physically-realistic general circulation model (GCM). In the latter, the narrow equatorial jet driven by the BD's rapid rotation manifests as an eclipse depth asymmetry, which our model reproduces as a latitude-dependent phase offset. Such a model, however, is only statistically preferred at approximately $3 \times$ the signal-to-noise ratio of the JWST data. When fitting to the actual JWST data of WD 1202, the full-phase eigenbasis model is statistically preferred. However, the resulting model map of WD 1202 B does not resemble the GCM predictions, with a concentrated day-side hotspot at the sub-stellar point with an angular size $\approx 15^\circ$. Interpreting this result more carefully requires follow-up work to disentangle systematics from true emission signals. This model framework provides a means to detect emission features from these rare types of eclipses, and ultimately to test predictions of models of atmospheric circulation across hot Jupiters and their BD analogs.

\end{abstract}


\section{Introduction}
\label{sec:introduction}
With over 20 brown dwarfs known to orbit white dwarfs, BD--WD systems are increasingly the target of observations that aim to probe the evolution of both stellar and planetary systems. These binaries span a wide range of orbital separations, from wide $\sim 100$ AU orbits \citep[e.g.][]{Becklin1988,Bergeron1990,Steele2009,Casewell2024a} to very close orbits of $\lesssim 0.01$ au. One pathway to form close-in BD--WD systems arises when the primary evolves into a giant star and subsumes the secondary into its envelope. The envelope is subsequently ejected, leaving behind a newly-formed WD host to a sub-stellar companion whose orbit can reach periods as short as $\sim 1$--2 hours \citep[see e.g.][]{Maxted2006,Beuermann2013,Steele2013,Parsons2017,Rappaport2017,Casewell2015,Casewell2018b,Casewell2018a,Nelson2018}. The instellations these companions receive resemble those of ultra-hot Jupiters, but with orbital periods --- and, assuming these companions are likely to be tidally locked --- rotation periods much shorter than HJs ($\sim 1$--a few days). Their periods are shorter even than many young ($\sim 1$--100 Myr) planetary-mass companions (PMCs), whose rotation periods are typically of order a few--10 hours \citep[e.g.][]{Lew2016,Manjavacas2019b,Zhou2019,Zhou2020,Landman2024}. Works such as \citet{Casewell2018a, Zhou2022a} and \citet{Amaro2025} show that for irradiation fluxes $\gtrsim 10^9$ erg cm$^{-2}$ s$^{-1}$, there are significant day-night contrasts, with the eclipsing system EPIC 212235321 (hereafter referred to as EPIC 2122) showing a rotational flux semi-amplitude of $\sim 30$\%. Subsequent 1-D and 3-D modeling of this system predicts a spectrally distinct hotspot and an equatorially concentrated feature of width $\approx 15^\circ$ \citep{tanAtmosphericCirculationUltrahot2019,Lothringer2020,Lothringer2024}. It is also for these extremely rapid rotators that we expect a more latitudinally-concentrated version of the super-rotation that is a hallmark in models of HJ atmospheres \citep[see e.g.][]{Showman2010b,Showman2011}. The Rossby numbers become small except at very low latitudes, and the deformation radius shrinks accordingly, producing a narrow ($\sim 10^\circ$) equatorial jet with peak eastward velocity at several km s$^{-1}$ \citep[see e.g.~][]{Tan2020}. The presence of an equatorial jet has to date not been detected observationally in any of these systems, and the predicted narrow latitude range weakens the possible signal for a day-side phase offset, consistent with the lack of detection of a significant offset in orbital phase curves of close-in BD--WD systems. Therefore, we lack evidence of an equatorial jet from observations to date \citep[e.g.][]{Amaro2025}.

BD companions around WDs have observational advantages that make them favorable testbeds for theories of sub-stellar atmospheric circulation. In terms of photometry, the spectral energy distributions (SEDs) of the WD and BD are highly distinct, particularly in the infrared where the BD's SED is able to stand out over the Rayleigh-Jeans tail of the WD's SED. This contrast is aided by the WD's uniform surface emission, which lacks the starspots or convective faculae that complicate observations of main-sequence hosts. In terms of dynamics, pre-CV BD--WD binaries typically have neither ongoing mass transfer nor significant tidal distortion, and the ultra-short orbital periods enable one to observe full-orbit phase curves with a high signal in just $\sim 1$ hour. The close-in orbits also make for a high probability of observing the system in eclipse; however, because the WD radius is intrinsically smaller than that of the BD, the secondary eclipses of the BD by the WD are strictly partial. The WD 1202-024 system (hereafter WD 1202) exhibits all these properties \citep{Eisenstein2006,Rappaport2017}, and WD 1202 B is expected to host similar extreme atmospheric structures to EPIC 2122 B. It therefore serves as a promising basis for our eclipse mapping model.

Eclipsing systems allow the spatial distribution of the emission across the companion disk to be probed with eclipse mapping \citep[as put forth in works such as][]{Horne1985,Williams2006,Rauscher2007b}. While phase curve observations allow one to detect offsets between the peak brightness of the day side and the sub-stellar point --- such as an eastward longitudinal shift from atmospheric super-rotation --- eclipse mapping unlocks the potential for constraining latitudinal structures. Recent eclipse mapping analyses with JWST data\footnote{Note that JWST will not be the only space observatory to be capable of eclipse mapping. Works such as \citet{Valentine2025} demonstrate the capability of the Ariel mission to perform a mapping survey of $\sim 100$ targets.} demonstrate evidence for latitudinal asymmetries, such as in WASP-43 b \citep{Challener2024,hammondTwodimensionalEclipseMapping2024}, and evidence for spectral gradients due to temperature and chemistry gradients as one moves away from the sub-stellar point \citep{Challener2025}. In principle, these techniques can be applied to systems like WD 1202, using the secondary eclipse to empirically isolate the day-night contrast --- whose contribution dominates the signal in the phase curve --- from smaller-scale latitudinal emission features. The partial eclipse of the BD by the WD, depending on the impact parameter, can occult at a similarly narrow range of latitudes to where a super-rotating jet would advect flux eastward. This would impart a phase offset whose light curve signal would be most significant \emph{in eclipse}; therefore, this is one example where eclipse mapping has the potential to recover a physically meaningful signal from the atmospheric circulation that is invisible in the phase curve. As long as one can precisely constrain the system's orbital geometry and the relative sizes of the bodies, we have the foundation for solving the inverse problem of retrieving a 2-dimensional spatial map from a 1-dimensional time series.

The full occultation of close-in giant exoplanets allows for a map fit with eclipse data that is continuous across the entire planetary day side \citep[see e.g.][]{Majeau2012,deWit2012,Demory2016,Challener2024,Challener2025,Lally2025}. However, because the secondary eclipses in BD--WD systems are inherently partial, we acquire additional spatial information for only a portion of the companion's day side. We must therefore make a decision in how our model stitches together the information in the eclipsed and non-eclipsed regions on the BD. In this work we develop a generalized model framework for mapping BD companions to WDs that quantifies and then provides one solution to this geometric limitation. The framework comprises two model components that are described in Section \ref{sec:model_methods}. Each relies on an initial orbital fit to the non-eclipse data of WD 1202 to constrain the orbital geometry of the system. We then test our model by fitting to simulated light curves of two cases where the BD exhibits only latitudinal structures (i.e.~zero phase variations), in Section \ref{sec:model_testing_in_latitude}; and to a light curve of a more realistic emission map derived from a previous climate model fit to a non-eclipsing BD--WD system in Section \ref{sec:GCM_test_results}. We then apply the model framework to the wavelength-integrated JWST NIRSpec light curve of WD 1202, and discuss the limitations of the model with our current instrumental precision (Section \ref{sec:discussion} and \ref{sec:conclusions}).

\section{The WD 1202 System and Predictions for BD--WD Eclipse Mapping}
\label{sec:WD1202}
\subsection{System Properties}
\label{sec:WD1202:system_properties}
WD 1202 is a pre-CV system with a mutual orbital period of approximately 71 minutes \citep{Parsons2017}, with the secondary WD 1202 B just below main-sequence mass \citep[$\sim \! 0.06 M_\odot$, see][]{Rappaport2017}. WD 1202 A has one of the hottest effective temperatures \citep[$\sim \! 22000$--24000 K, see][]{Eisenstein2006,Kleinman2013,Parsons2017}, with \citet{Rappaport2017} reporting a mean day-side temperature of $4390\pm375$ K for WD 1202 B and an upper limit of $\lesssim 2450$ K for the night side. There is neither significant evidence for tidal distortion of the companion, nor for any mass transfer onto WD 1202; WD 1202 B is believed to be well under-filling its Roche lobe \citep[by $\sim 3$--17\%, as originally reported in][]{Rappaport2017}. We can therefore assume the projected disk of WD 1202 B is approximately circular for the purposes of eclipse mapping.

\subsection{Observations and Data Reduction}
\label{sec:WD1202:observations}
WD 1202 was observed across a full orbital period as part of a JWST observing program (GO Program ID 4967; PI: Yifan Zhou) with NIRSpec PRISM/CLEAR ($R \sim 30$--300, 0.6--5.3 $\upmu$m) in the Bright Object Time-Series (BOTS) mode. The observation covers 90.7 minutes across 50 integrations of 120 groups each. The WD 1202 NIRSpec/PRISM observations were reduced using the JWST calibration pipeline v1.18.0 \citep{Bushouse2025}, CRDS context \texttt{jwst\_1371.pmap}, and a customized extraction based on \texttt{Eureka!} v1.2 \citep{Bell2022}. After calibrating the full detector ramps with an $8\sigma$ jump-detection threshold, we divided each of the 50 integrations into 12 non-overlapping segments of ten groups, retaining the four frames per group. The pipeline's \texttt{OLS\_C} ramp-fitting algorithm yielded 600 time-resolved images with nominal integration times of 9.0464~s. Barycentric mid-exposure times were calculated from the original integration midpoints and the mean group-index offsets. Stage~2 provided the wavelength calibration without flat-field, path-loss, or photometric corrections, preserving the native detector sampling. For each image, a constant background was estimated independently in each detector column from unmasked rows 8--10 and 21, with $7\sigma$ temporal and $3\sigma$ spatial rejection. We extracted the source within rows 12--18 (zero-based index) using a generalized optimal extraction \citep{Horne1986} and the \texttt{Eureka!} median-image spatial profile with a 13-column smoothing window. The covariance matrix $\mathbf{C}=\mathrm{diag}(V_y)+V_B\mathbf{1}\mathbf{1}^{\mathsf T}$ accounted for the background error shared by source pixels within each column, where $V_y$ is the detector variance, $V_B$ is the variance of the background mean, and $\mathbf{1}$ is a vector of ones matching the length of the diagonal. For a normalized spatial profile $\mathbf{P}$ and background-subtracted data $\mathbf{D}$, the extracted signal and variance were $S=(\mathbf{P}^{\mathsf T}\mathbf{C}^{-1}\mathbf{D})/(\mathbf{P}^{\mathsf T}\mathbf{C}^{-1}\mathbf{P})$ and $\sigma_S^2=(\mathbf{P}^{\mathsf T}\mathbf{C}^{-1}\mathbf{P})^{-1}$. Each retained column required at least three valid aperture pixels and at least 50\% of the spatial-profile support.

The white-light curve was constructed by summing the 393 columns with valid support at all times over $0.6\leq\lambda<5.3\,\upmu$m. Its variance was obtained by summing the corresponding column variances. These measurements are in electron rates and no wavelength-bin-width weighting was applied. The extracted light curve showed a repeatable pattern across the twelve subintegration positions within each parent integration. To isolate this instrumental pattern from the astrophysical variability, we represented the orbital phase curve with a constant and four Fourier harmonics at a period of 71.23~min. An empirical multiplicative correction for each subintegration position, containing twelve independent factors that averaged to 1, was fitted jointly with the phase curve. The fit excluded the first parent integration and samples within 0.055 orbital phase of either eclipse center. We validated the correction by dividing alternate parent integrations into two subsets and correcting each subset using the pattern inferred from the other. This procedure preserved the astrophysical phase variation while suppressing the systematic signal near the 108.8~s parent-integration cadence. The phase-curve residual RMS decreased from approximately 6460 to 5290~ppm, with the corrected scatter about 1.3 times the nominal statistical uncertainty. Both the flux and its statistical uncertainty received the same multiplicative correction and were normalized by the median uncorrected rate over the fitting sample ($\approx 13589.74$ $e^{-}$ s$^{-1}$). The final light curve retains all 600 samples without temporal smoothing or empirical uncertainty rescaling. The fitted correction covariance was retained separately for subsequent uncertainty propagation.

The typical signal-to-noise ratio (SNR) of the flux relative to the per-point error bar is $\sim 7$ within the time where the WD is fully occulted (primary transit), and increases to $\sim 245$ just outside. The wavelength-integrated eclipse depth corresponds to a signal $\approx 6$ times the typical estimated per-point uncertainty during eclipse. We use this value as a reference point for injection-and-recovery tests in Section \ref{sec:model_testing_in_latitude}--\ref{sec:GCM_test_results}.

\section{Modeling Methodology}
\label{sec:model_methods}
\subsection{General Methodology for BD--WD Eclipse Mapping}\label{sec:model_methods:eclipse_mapping}
To determine the time-resolved contribution of the companion's emission in latitude and longitude, we fit for the orbital period, mass and line-of-sight orbital inclination of the host, and the radii of both bodies. We use all data except those in secondary eclipse, i.e.~the phase curve and primary transit, and assume Keplerian orbits. A previous study \citep{Zhou2022a} on HST data of EPIC 2122, a non-eclipsing BD--WD system \citep[first reported in][]{Casewell2018a}, used a Fourier series model in time for the phase curve data that adequately reproduced the rotational modulations in the light curve due to the day-night contrast. EPIC 2122 B is similar to WD 1202 B in its ultra-short $\sim 1$ hr orbital period and $\gtrsim 4000$ K day-side temperature. To constrain the orbital and transit geometry of WD 1202 with the broad-band 0.6--5.3 $\upmu$m data, we adopt a similar model, applying zonal spherical harmonics to the emission map of the BD to produce the sinusoid harmonics in the light curve.\footnote{Here, ``zonal'' assumes the polar axis originates from the sub-stellar point, such that the spherical harmonic dipole ($Y_{1, 0}$) is aligned with the day-side and night-side hemispheres.}. We then present two complementary approaches for modeling a BD--WD system light curve with our constrained viewing geometry. The first approach applies the eigenmapping technique to our partially-occulting case, which we will show produces maps that are effectively super-positions of ``phase-like'' and ``eclipse-like'' structures (Section \ref{sec:model_methods:eigenmapping}). This technique also provides a straightforward method to select the appropriate complexity of a model that avoids over- or under-fitting. We then use the results of our model selection with the eigenmapping-derived basis of light curves to tune a second type of model (detailed in Section \ref{sec:model_methods:complexity_penalization}). This model instead prioritizes fitting the light curves with the simplest spherical harmonic degrees that yield a solution of equal goodness of fit.

\subsection{Constraining the Viewing Geometry for Eclipse Mapping with an Orbital and Phase Curve Fit}\label{sec:model_methods:orbital_fit}
We perform our initial phase curve fit on the broad-band 0.6--5.3 $\upmu$m data of WD 1202 with secondary eclipse removed (approximated as $t= 60804.46074$--60804.46798 BMJD$_\mathrm{TDB}$). The free parameters are listed in Table \ref{table:orbital_fit_results} and include the orbital period, time of mid-transit, the host mass, radii and mean emission fluxes of both bodies, and the impact parameter of the WD as it transits in front of the BD. The WD's emission is treated as uniform except for limb darkening, which is modeled to second order \citep[see][]{Kipping2013}.\footnote{WD-specific limb-darkening coefficients are well established at optical wavelengths \citep{Gianninas2013,Claret2020}, but have not been computed for near-infrared bandpasses; existing near-infrared coefficients are restricted to lower surface gravities \citep{Claret2000a}. Limb darkening is expected to be weak at NIRSpec wavelengths, so we adopt the existing quadratic limb-darkening law in \texttt{jaxoplanet}.} For the BD we use a model that includes zonal spherical harmonic modes up to and including degree $\ell=6$. We follow the convention in \texttt{jaxoplanet} with the polar axis at the sub-stellar point.\footnote{Beyond the dipole term, only even wavenumbers (equivalently, even-degree zonal harmonics) produce phase variations in rotational light curves; see Section \ref{sec:model_methods:eigenmapping} and \citet{cowanInvertingPhaseFunctions2008}, \citet{cow13}. In our tests, including odd-degree zonal harmonics, even though they are effectively invisible in the phase curve, allows the model to find a solution whose map avoids negative fluxes.} We extend our wavenumbers to 6th order in anticipation that this will be just beyond the complexity needed to fit our phase curve. Our final free parameter is an error inflation parameter that multiplies all error bars on the data by a uniform factor.

Unlike mapping studies with eclipsing light curves of hot Jupiters, such as those of WASP-43 b \citep{hammondTwodimensionalEclipseMapping2024,Challener2024}, we do not include instrumental effects directly in our model, such as an exponential ramp \citep[see e.g.][]{Bell2024} or a linear trend \citep[see e.g.][]{schlawinPlanetEclipseMapping2023}. Such drifts are seen in other JWST time-series modes, with MIRI LRS observations of WASP-43 b showing a $\sim$0.1\% exponential ramp over $\sim 1$ hour combined with a $-39$ ppm hr$^{-1}$ linear trend \citep{Bell2024}. A study on NIRSpec BOTS data by \citet{Espinoza2023} required only the linear component. For the linear trend, \citet{Espinoza2023} and the JWST User Documentation\footnote{\url{https://jwst-docs.stsci.edu/known-issues/nirspec-known-issues/nirspec-bots-known-issues}} report $\sim140$ ppm hr$^{-1}$ for NRS1 observations of HAT-P-14 b, with no ramp component.\footnote{No direct estimate exists for NIRSpec PRISM, although \citet{Rustamkulov2023} removed a constant 200 ppm hr$^{-1}$ linear trend from their PRISM observations of WASP-39 b.} There is no clear exponential trend at the start of the exposure in the residuals of our orbital phase curve fit, and a linear fit to the residuals returns a slope of $\approx+1.9$ $e^{-}$\,s$^{-1}$\,hr$^{-1}$, or $\approx 145$ ppm hr$^{-1}$ relative to the median phase curve flux. At only $\sim 3$ e$^{-}$\,s$^{-1}$ accumulated over the observation, this is well below our estimated noise floor (see below).

We use \texttt{jaxoplanet}, a multi-purpose exoplanet orbital and transit fitting tool, to generate the model light curves. Our light curves are evaluated at a time resolution 100 times finer than the data, then binned to the data time sampling using \texttt{spectres} \citep{Carnall2017}. The model parameter optimization is performed using \texttt{Nautilus} \citep{Nautilus}, a Python package that uses a combination of nested sampling with a neural network to propose efficient bounds for sampling. We use normal priors for the BD orbital period, WD mass, and radii, tuned to constraints reported in \citet{Parsons2017} and \citet{Rappaport2017}. The priors on the remaining parameters are kept uniform and broad; all prior distributions are listed in Table \ref{table:orbital_fit_results}. We initiate the sampler with 1000 live points, use the default stopping condition of 10,000 effective samples, and discard the exploration stage that precedes the estimate of the posterior distributions.\footnote{See the \href{https://nautilus-sampler.readthedocs.io/en/latest/}{\texttt{Nautilus} documentation}.}

\begin{figure*}
\begin{center}
\includegraphics[width=17cm]{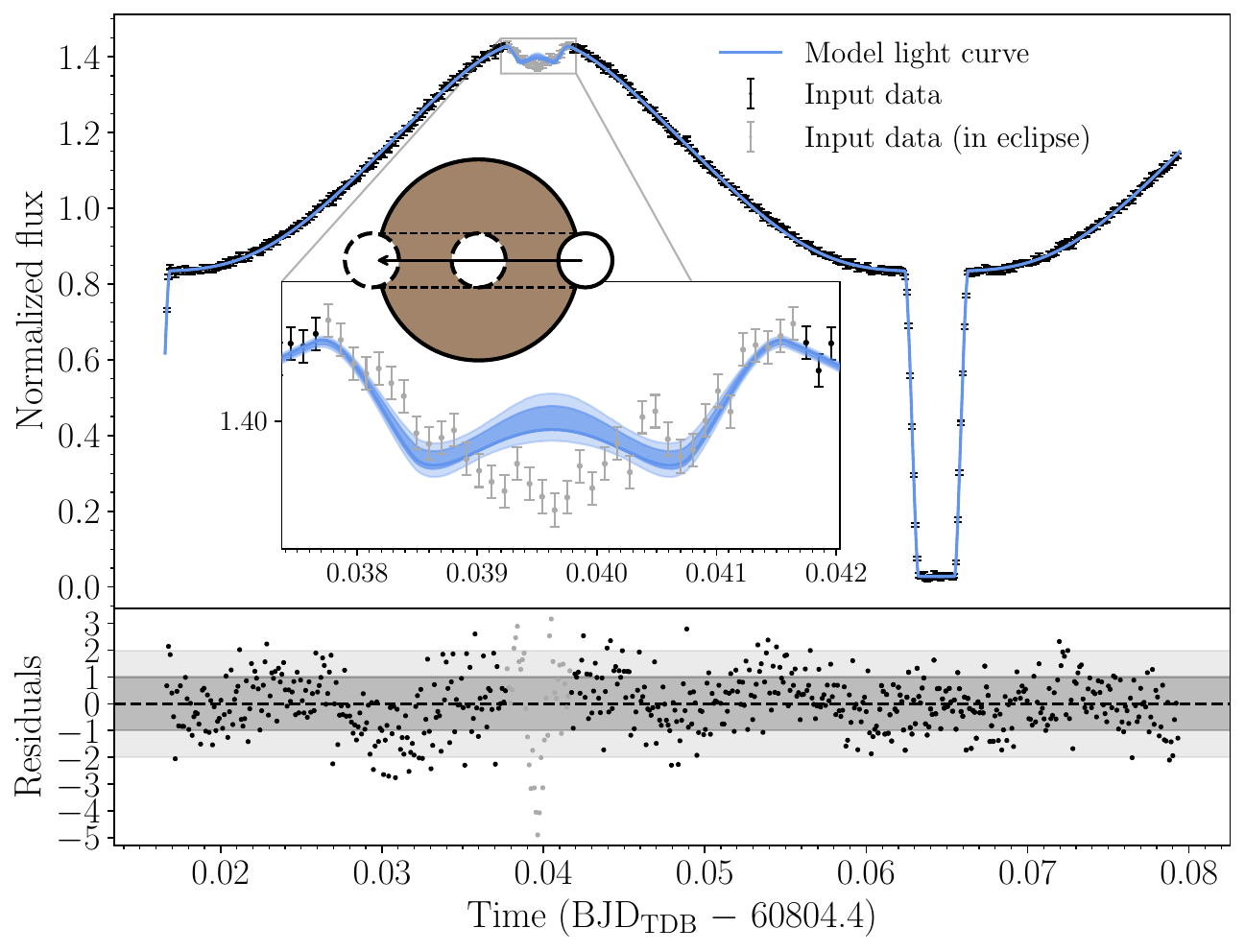}
\caption{The best-fit light curve (blue) to the phase curve and transit data of WD 1202 (black points with error bars), integrated across 0.6--5.3 $\upmu$m. The data have been normalized to a reference median flux count of $\approx 13589.74$ $e^{-}$ s$^{-1}$. The inset light curve plot shows a zoom-in to secondary eclipse, with the data excluded from the initial phase curve fit in grey. The shades of blue denote the 68th and 95th percentiles of the extrapolated light curves in eclipse. On top of the inset is a diagram showing a possible secondary eclipse geometry for the system, using the maximum-likelihood estimate for the impact parameter and relative radii. The bottom panel shows the residuals scaled to the model uncertainties, which have been inflated by a constant factor of $\approx 1.4$ from those calculated in the reduction. The error inflation factor is treated as a free parameter in our model.}
\label{fig:orbital_fit}
\end{center}
\end{figure*}

Our results are reported along with the priors in Table \ref{table:orbital_fit_results}; the fit is shown in Figure \ref{fig:orbital_fit}. Our data are too coarsely-sampled in time to provide a more precise constraint to the orbital period and transit time than those in the literature. The posterior distribution for the period closely resembles the prior, and we retrieve an $\approx  \pm 120$ ms uncertainty for the time of mid-transit. Our results for the masses and radii are both consistent with the priors. Our inclination is constrained to a 68\% credible interval of $88.^\circ1^{+1.2}_{-1.0}$. A degree of uncertainty of inclination, however, corresponds to a wide range of plausible impact parameters for the WD relative to the BD: $b=0.64^{+0.34}_{-0.40}$, with a maximum-likehood estimate (MLE) of $b \approx 0.015$. This relatively large uncertainty in impact parameter arises from the peculiarity of the geometry of secondary eclipses in cases where the host is smaller in radius than the companion. The wings of the light curve in eclipse, i.e.~during ingress and egress, are shaped by the impact parameter. At $b=0$, the host will move into the companion's projected disk at an angle normal to the edge of the disk. As the impact parameter increases, that angle becomes increasingly shallow. Our results imply the change in angle is too small to impart a statistically significant change to the shape of the wings out to $b \approx 1$.\footnote{Note also that the impact parameter is defined relative to the WD, which has a smaller radius than the BD, so $b$ can exceed 1.} The large range also implies we will not have a precise constraint on the latitude range that the WD eclipses on the BD. For the purposes of our tests in Section \ref{sec:model_testing_in_latitude}, where only latitudinal structure is simulated, we use the median value of the impact parameter from its posterior: $b_\mathrm{med} \approx 0.64$. We choose the median because a significantly non-zero impact parameter is needed to produce a non-zero signal in the dipole test (Section \ref{sec:model_testing_in_latitude:dipole_test}). For the fit to the simulated GCM data (Section \ref{sec:GCM_test_results}), as well as to the real data (Section \ref{sec:white_light_fit}), we use the maximum-likelihood estimate $b_\mathrm{MLE} \approx 0.015$. The WD then transits in front of the BD nearly centrally, in an approximately symmetric latitude range of $\approx 16^\circ$S--$16^\circ$N; at this near-central geometry neither hemisphere is preferentially occulted.

We are able to constrain the spherical harmonic terms of the BD phase curve out to $Y_{5, 0}$ without secondary eclipse data; the $Y_{6, 0}$ term is consistent with zero. Note that we do not fit to the data in secondary eclipse; the structure seen in the resulting models during secondary eclipse is indicative of a day-side ``hotspot'', a feature that these harmonics permit and which we will see in a similar form when discussing the ``spot and dipole'' models in Sections \ref{sec:GCM_test_results} and \ref{sec:white_light_fit}. The model is unable to find a solution to the remaining data (phase curve and primary transit) with a reduced chi-squared metric close to 1 without inflating the error bars by a factor of $\approx 1.4$. The residuals do not show any visible trends, with the exception of slightly increased residuals within the ingress and egress of primary transit. Figure \ref{fig:orbital_fit} shows the extrapolation of the model light curves within eclipse, with the 68- and 95\%-ile range of fluxes from re-sampling 3,000 draws from the posterior. The inflated error bars slightly exceed the 95\% range of fluxes. We adopt the inflated error bars from our maximum likelihood phase curve model for our fit to the near-infrared light curve of WD 1202 in Section \ref{sec:white_light_fit}, as well as for the test cases on simulated data (with additional variants at the uninflated errors and at higher SNR for the GCM test cases, described further in Section \ref{sec:GCM_test_results}).

\begin{deluxetable*}{r|cccc}
\tablecaption{Keplerian Orbital Fit Result for WD 1202}
\tablehead{
\colhead{Parameter (Units)} & Prior & \colhead{Median $\pm 1\sigma$} & \colhead{Maximum Likelihood Estimate (MLE)}
}
\startdata
$P_\mathrm{orb}$ (days)     & $\mathcal{N}\!\left(0.049465251\,\mathrm{d}, 0.5\,\mathrm{ms}\right)$ & $0.04946525$ d ${}_{-0.5}^{+0.5}$ ms & 0.04946525 \\
$t_0$                       & $\mathcal{U}\!\left(t_\mathrm{obs, start}, t_\mathrm{obs, end}\right)$   & ${\bar{t}_0}{}_{-118.4}^{+116.2}$ ms       & $\bar{t}_0 - 10.8$ ms \\
$M_{\text{WD}}$ (M$_{\odot}$) & $\mathcal{N}\!\left(0.40, 0.02\right)$    & $0.412 \pm 0.015$     & 0.405 \\
$R_{\text{WD}}$ (R$_{\odot}$) & $\mathcal{N}\!\left(0.0219, 0.0003\right)$    & $0.0218 \pm 0.0003$           & 0.0220 \\
$R_{\text{BD}}$ (R$_{\odot}$) & $\mathcal{N}\!\left(0.085, 0.003\right)$ & $0.0831_{-0.0014}^{+0.0017}$             & 0.0814 \\
$b_{\text{WD}}$             & $\mathcal{U}\!\left(0, R_\mathrm{WD}/R_\mathrm{BD} + 1\right)$  & $0.64_{-0.40}^{+0.34}$        & 0.015 \\
$i_{\text{WD}}$ (${}^\circ$)     & N/A & $88.1_{-1.0}^{+1.2}$        & 90.0 \\
$F_{\text{WD}}$ (normalized)     & $\mathcal{U}\!\left(0.5, 1.5\right)$ & $0.805 \pm 0.002$    & 0.806 \\
$F_{\text{BD, mean}}$ (normalized) & $\mathcal{U}\!\left(0, 1\right)$  & $0.263 \pm 0.002$      & 0.261 \\
BD $Y_{1,0}$                & $\mathcal{U}\!\left(-2, 2\right)$ & $1.009 \pm 0.006$ & 1.013 \\
BD $Y_{2,0}$                & $\mathcal{U}\!\left(-2, 2\right)$ & $0.428 \pm 0.005$ & 0.431 \\
BD $Y_{3,0}$                & $\mathcal{U}\!\left(-2, 2\right)$ & $-0.135_{-0.046}^{+0.045}$ & $-0.080$ \\
BD $Y_{4,0}$                & $\mathcal{U}\!\left(-2, 2\right)$ & $-0.261 \pm 0.024$ & $-0.264$ \\
BD $Y_{5,0}$                & $\mathcal{U}\!\left(-2, 2\right)$ & $-0.147_{-0.050}^{+0.045}$ & $-0.119$ \\
BD $Y_{6,0}$                & $\mathcal{U}\!\left(-2, 2\right)$ & $-0.007_{-0.043}^{+0.038}$ & 0.025 \\
$q_0$                        & $\mathcal{U}\!\left(0, 1\right)$        & $0.48_{-0.26}^{+0.31}$      & 0.54 \\
$q_1$                       & $\mathcal{U}\!\left(0, 1\right)$        & $0.30_{-0.20}^{+0.33}$       & 0.10 \\
$u_0$                       & N/A        & $0.39_{-0.26}^{+0.32}$      & 0.15 \\
$u_1$                       & N/A        & $0.27_{-0.40}^{+0.38}$      & 0.58 \\
$\log_{10} f_{\text{err}}$  & $\mathcal{N}\!\left(-3, 3\right)$ & $0.157 \pm 0.013$ & 0.146
\enddata
\tablecomments{$\pm 1 \sigma$ bounds are calculated from the 68\% credible interval of our fit. The weighted mean time of mid-transit is $\bar{t_0} = 60804.46435801$ BMJD$_\mathrm{TDB}$. Our normal priors are drawn from taking the results reported in \citet{Parsons2017} and \citet{Rappaport2017} and merging the normal distributions constructed from their best-fit parameter values and uncertainty ranges. Fluxes and spherical-harmonic amplitudes are expressed in the same normalized units as the light curve. The upper bound for our prior on the impact parameter is determined by the ratio of the means of the merged radius priors. All parameters listed with ``N/A'' as a prior are not free, but instead derived from free parameters, including the orbital inclination and the limb-darkening parameters via $u_0 = \sqrt{q_0} \left(2 q_1\right)$ and $u_1 = \sqrt{q_0} \left(1 - 2 q_1\right)$, as in \citet{Kipping2013}.}
\label{table:orbital_fit_results}
\end{deluxetable*}

\subsection{Eigenmapping Methodology}\label{sec:model_methods:eigenmapping}
\emph{Eigenmapping} refers to one approach to eclipse mapping where a basis of light curves is generated using the specific geometry of the eclipsing system \citep[e.g.][]{Rauscher2018,Mansfield2020,Adams2023a,challenerThERESAThreeDimensionalEclipse2022,challenerEclipsemappingNullSpace2023,Challener2024,Challener2025}. An eigenmapping model distills the additional empirical information available from observing the companion in secondary eclipse, particularly information in latitude \citep[see e.g.][]{cow13,Cowan2017,challenerEclipsemappingNullSpace2023}. Since we have an incomplete description of the 2-dimensional emission map from our light curve, a 1-dimensional Fourier series is no longer a sufficient model, and a fully 2-dimensional basis such as spherical harmonics is under-constrained. Eigenmapping constructs one solution that is tuned to empirically characterize which structures in two dimensions our light curves can measure. To construct our new basis, we employ the theory of Slepian functions (\citealp{Slepian1964}, with a more recent overview in e.g.~\citealp{Simons2010}, Section 2.3). In the same way spherical harmonics are solutions to Laplace's equation on a full spherical surface, we can construct a set of orthogonal basis functions that are constrained by our limitation in sampling the companion's emission in both space and time. Since any flux $F\!\left(\theta, \phi\right)$ on a spherical body can be described with spherical harmonics, we may express our new \emph{eigenfunctions} as linear combinations of spherical harmonics. These eigenfunctions distill the sub-set of harmonic space that produces observable signals in our light curves. To construct these functions, we maximize the amount of \emph{power} $\lambda$ in the observed light curve of the emitted flux:
\begin{equation}\label{eq:slepian_constraint}
    \max \lambda = \frac{\int_{\mathcal{T}} L^2\!\left(t\right) \, dt}{\int_\Omega F^2\!\left(\theta, \phi\right) \, d\Omega}
\end{equation}
where $\Omega$ denotes the surface of the sphere, and $\mathcal{T}$ the observation time. The observed light curve from a given body in the system is
\begin{equation}\label{eq:lightcurve_integral}
    L\!\left(t\right) \equiv \int_\Omega F\!\left(\theta, \phi\right) V\!\left(\theta, \phi; t\right) \, d\Omega.
\end{equation}
$V\!\left(\theta, \phi; t\right)$ is the function that describes which regions of the companion's photosphere are visible at a given time \citep[see e.g.~][]{cow13,Luger2019,Agol2020}. We construct a basis $L_{\ell, m}\!\left(t\right)$ of light curves generated by spherical harmonic maps, and express the light curve $L$ produced by a map $F$ with the same linear combination of harmonic coefficients as $F$. By building the model from this basis of light curves rather than the underlying spherical harmonics, eigenmapping directly incorporates the observational limits in space and time.

We solve Equation \ref{eq:slepian_constraint} using singular value decomposition \citep[for previous applications, see][]{lugerMappingStellarSurfaces2021, challenerEclipsemappingNullSpace2023}, implemented with the principal component analysis \citep[PCA, as originally described in][]{Hotelling1936} routine in \texttt{scikit-learn} \citep{scikit-learn}. PCA takes a set of basis vectors of some data and transforms the basis to a new set of vectors that are ordered by variance. Here the variance is the signal in the light curve, and the data are the basis $\matnot{H}_{n, t}$ of pure spherical harmonic light curves (indexed by $n \equiv \ell \left(\ell+1\right) + m$). We would like to find $\matnot{\eta}_{i, n}$, the matrix that maps the necessary coefficients of a linear combination of harmonic light curves to each new eigencurve $\matnot{E}_i$, the eigenvectors of the new basis: $\matnot{E}_{i, t} = \matnot{\eta}_{i, n} \matnot{H}_{n, t}$. We solve for $\matnot{\eta}$ by maximizing
\begin{equation}\label{eq:PCA_slepian_constraint}
    \max \lambda = \frac{1}{N_t-1} \max \left\{ \frac{\| \matnot{\eta}_{i, n} \matnot{H}_{n, t} \|^2}{\|\matnot{\eta}_{i, n} \|^2} \right\}
\end{equation}
where $N_t$ is the number of time elements. This is done successively in $i$ with the condition that each new eigenmode must be orthogonal to all previous.

Once an eigenbasis is constructed, the fitting process to data is a straightforward linear algebra problem of minimizing
\begin{equation}\label{eq:eigenbasis_fit}
    \min \left\| \matnot{E}_{t, i} \matnot{e}_i -  \matnot{d}_t \right\|^2
\end{equation}
which we perform using the weighted least-squares \texttt{WLS} routine in the \texttt{statsmodels} package \citep{statsmodels}. Here $\matnot{E}_{t, i} = \matnot{E}^{\mathsf T}_{i, t}$, $\matnot{e}_i$ are the coefficients of the fit to the eigenbasis, and $\matnot{d}_t$ are the data. We weight the fit according to the inverse of the squared uncertainties in the data at each time. It is important to note that while the eigenbasis vectors are designed to be mutually orthogonal, with uncertainty weighting there will be non-negligible covariances among the fit parameters. As with the fit to the transit data, we generate our model light curves at $100 \times$ the time resolution of the JWST data (60,000 versus 600). We use \texttt{jaxoplanet} to generate the initial basis of spherical harmonic light curves, and the \texttt{pca} routine in \texttt{scikit-learn} \citep{scikit-learn} to evaluate $\matnot{\eta}_{i, n}$ and the resulting set of eigencurves $\matnot{E}_{i, t}$. We use \texttt{jaxoplanet} to evaluate the eigenmaps that produce each eigencurve on a latitude/longitude grid.

\begin{figure*}
\begin{center}
\includegraphics[width=17cm]{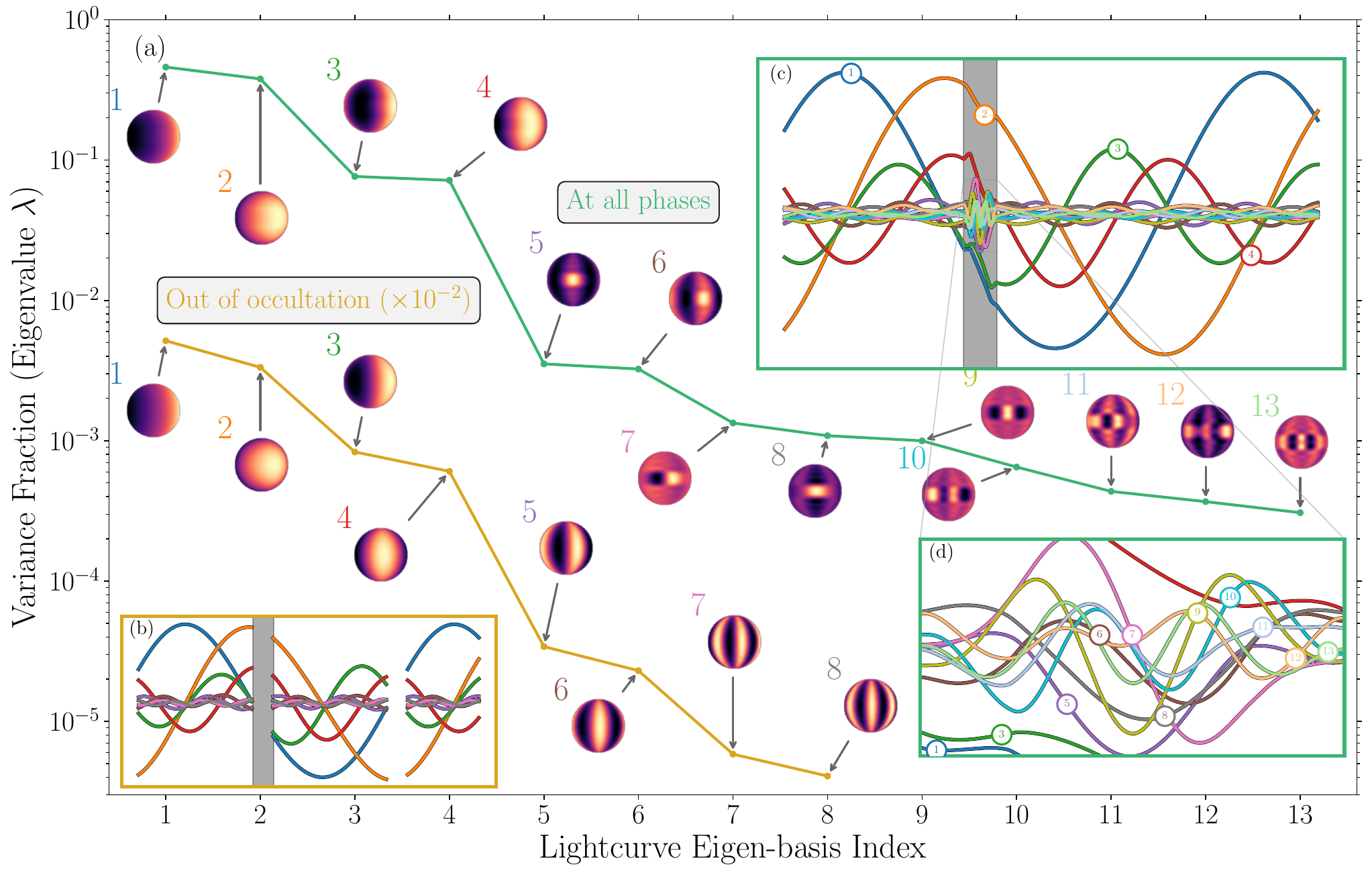}
\caption{The first several components of the phase-only and full-phase eigenbases for WD 1202. The main plot shows the fraction of variance explained as a function of the index of the eigenbasis component (eigenmode); this is equivalent to the eigenvalue (see Section \ref{sec:model_methods:eigenmapping}). The phase-only variance fractions are offset by two orders of magnitude for visual clarity. Inset plot (b) shows the light curves from each pure eigenmode for the phase-only eigenbasis, with colors corresponding to the numbers on the variance plot. Insets (c)--(d) show the corresponding light curves for the full-phase eigenbasis, with the grey shaded region denoting the time interval of secondary eclipse. Inset (d) shows the structure of the eigencurves within secondary eclipse (grey shading removed for clarity). The first 4 components (``eigenmodes'') appear similar to those of the phase-only eigenmodes, indicating most of the signal remains in the rotation at low harmonic degrees. Starting at the 5th eigenmode and higher, we see a concentration of brightness fluctuations along the path of occultation by the WD host. These modes are super-positions of phase-like and eclipse-like modes. As we approach the higher modes, the eclipse signal begins to dominate as brightness variations in the occulted region are able to produce more variance (signal) in the light curve than including longitudinal bands at increasing wavenumbers.}
\label{fig:eigenvalues_and_maps_at_all_phases}
\end{center}
\end{figure*}

Figure \ref{fig:eigenvalues_and_maps_at_all_phases} shows eigenbases tuned to the WD 1202 system geometry. We generate the eigenbases for two cases: the ``phase-only'' data which excludes secondary eclipse and represents the information from rotation alone; and the ``full-phase'' data which includes secondary eclipse. The ``zeroth'' order for all bases is the constant map. While eigenmapping's key strength is fitting light curves with eclipses, we use the phase-only basis as a ``null hypothesis'' model with which to compare the fit to all the data (see Section \ref{sec:model_methods:model_selection}). The phase-only eigenmaps are dominated by longitudinal bands that produce sinusoidal light curves. The dipole ($\ell=1$) is the only odd harmonic degree that can contribute in the phase curve; all odd $\ell \geq 3$ zonal harmonics will have a constant hemispheric-integrated flux as the body rotates \citep[a known result, see e.g.~][]{cowanInvertingPhaseFunctions2008,cow13}. The full-phase eigenbasis is a super-position of phase-only maps with patterns of increasing wavenumber within the occultation region (Figure \ref{fig:eigenvalues_and_maps_at_all_phases}). The maximum signal within the occultation is weaker than that of the phase variations; we only reach an eigenmode with appreciable eclipse signal at the 5th eigenmode. Beyond this eigenmode, the amplitudes of variations within eclipse compete with those of the phase variations.

\subsection{Model Selection Criteria}\label{sec:model_methods:model_selection}
We evaluate the evidence for eclipse data providing additional information about the companion's emission by setting the model fit from the phase-only eigenbasis as our null hypothesis. That is, the fit with the full-phase basis must return a statistically significantly better fit than that of the phase-only basis.  An advantage of eigenmapping is that we can iterate in the number of free parameters simply by adding eigenmodes sequentially. Additionally, by construction, the covariance among the model light curves is minimized. With these properties of our model, a natural choice for model selection is to compare the Bayesian Information Criterion \citep[see e.g.~][]{Kass1995} of each model:
\begin{equation}\label{eq:BIC}
   \mathrm{BIC} = n_\mathrm{par} \log n_\mathrm{data} - 2 \log \mathcal{L}
\end{equation}
for a model with a number of free parameters $n_\mathrm{par}$ and fit to a number of data points $n_\mathrm{data}$. The log likelihood
\begin{equation}
    \log \mathcal{L} \equiv -\frac{1}{2} \left\{ \sum\limits_{t=0}^{n_\mathrm{t}} \left[\frac{F\!\left(t\right)-\hat{F}\!\left(t\right)}{\sigma\!\left(t\right)}\right]^2 + \log\left[2 \pi \sigma^2\!\left(t\right)\right] \right\}
\end{equation}
where $F$ is the model light curve, $\hat{F}$ the data, and $\sigma$ is the measurement uncertainty. The heuristic in \citet{Kass1995} prescribes that we prefer a lower BIC value, and that ``marginal'' evidence for one model against another occurs when that model's BIC is smaller than the other's by at least 2. A difference of 6 shows a ``moderate'' preference, and a difference of at least 10 is considered a ``strong'' preference. We compare first among fits using each eigenbasis to find the optimal number of eigenmodes both with and without eclipse, and then between the best fits to evaluate whether the full-phase model is preferred over the phase-only model. The latter step is excluded for the latitude-only tests in Section \ref{sec:model_testing_in_latitude} since these tests exclude phase variations.

\subsection{A Power Spectrum Prior for Global Map Estimation}\label{sec:model_methods:complexity_penalization}
The map generated from fitting eigencurves to a light curve is just one possible solution for an inherently under-constrained problem. Throughout this work we refer to such solutions as ``maps'', where they may more accurately be referred to as spatial models of an underlying ``true'' map. The models are estimates within a chosen basis and prior, with the components of the emission in the null space of the light curve \citep{challenerEclipsemappingNullSpace2023} set by those choices rather than by the data. Despite its usefulness for evaluating the evidence for information in eclipse --- and for selecting an appropriate model complexity --- our eigenmapping-derived solutions to the emission map typically introduce sharp gradients at the boundary of the occulted and un-occulted regions. This limits our ability to interpret the global map as a direct estimate of the BD's emission, as the emission is not fine-tuned to our view --- only our knowledge of it is. We can use our eigenmapping result to set a baseline metric for the quality of fit (e.g.~the log likelihood, or mean-squared error) that is statistically appropriate were one to use a non-eigenmapping model. One alternate approach is to prefer solutions whose map estimates vary only at the largest angular scales that still produce a fit of appropriate quality. While not directly a physical prior, we note that the BD receives irradiation from its host on a hemispherical scale, and therefore the thermal response of the BD will also respond to first-order on similarly large scales. We introduce a constraint to our matrix formulation that penalizes the power spectrum increasingly with spherical harmonic degree. This procedure is a form of \emph{Tikhonov regularization} \citep{Tikhonov1963}, itself a form of singular value decomposition.\footnote{For examples of general Tikhonov regularization, see Section 2.7.2 in \citealp{Bjorck1996}, Section 1.3 in \citealp{Hansen1998}, Section 2.3 in \citealp{Kaipio2005}, and Section 4.2 in \citealp{Aster2013}.} We start with our original eigenmapping constraint (Equation \ref{eq:eigenbasis_fit}) and add a regularization term

\begin{equation}\label{eq:square_linear_algebra}
    \min\left(\left\| \matnot{E}_{t, i} \matnot{e}_i -  \matnot{d}_t \right\|^2 + \alpha^2 \left\| \matnot{L}_{n,n} \matnot{c}_n \right\|^2\right)
\end{equation}
where $\matnot{c}_n$ are the spherical harmonic coefficients of our fit: $\matnot{e}_i = \matnot{\eta}_{i, n} \matnot{c}_n$. $\alpha$ is a free parameter that controls the strength of the penalization relative to the residuals of the initial fit. We construct $\matnot{L}$ using the eigenvalues of the Laplacian equation for spherical harmonics:
\begin{equation}\label{eq:penalty_exponent}
    \matnot{L}_{n, n}^2 = \left[\ell \left(\ell+1\right)\right]^p,
\end{equation}
where the degree $\ell$ at spherical harmonic index $n$ is $\lfloor \sqrt{n} \rfloor$ and, by default, $p = 2$. This is a \emph{second-order} Tikhonov regularization; one way to visualize this procedure is to apply a diffusion operator to the power spectrum, effectively blurring the angular resolution of the map.\footnote{Since works like \citet{Aster2013} use the context of a spatial basis, this means their regularization matrix $\matnot{L}$ has non-zero elements up to two spaces off the diagonal (also called a ``pentadiagonal'' matrix, as in Equation 4.28 of \citealp{Aster2013}). However, since we are operating in a spherical harmonic (spectral) basis, where we are using eigenfunctions of the Laplacian (second derivative) operator, our penalty matrix is purely diagonal with elements indexed by $\ell$.} Following Equations 4.25--4.26 in \citet{Aster2013}, we can rewrite this as an augmentation of $\matnot{E}$ with $\matnot{L}$:
\begin{equation}\label{eq:augmented_linear_algebra}
    \min\left(\left\| \begin{bmatrix} \matnot{E}_{t, i} \\ \alpha \matnot{L}_{n, n} \matnot{\eta}_{n, i} \end{bmatrix} \matnot{e}_i - \begin{bmatrix} \matnot{d}_t \\ \matnot{0} \end{bmatrix} \right\|^2\right)
\end{equation}
where $\matnot{\eta}_{n, i} = \matnot{\eta}^{\mathsf T}_{i, n}$.

For this model, the key additional free parameter is $\alpha$, which controls how strongly higher degrees are penalized. We note that while standard implementations of Tikhonov regularization typically determine the optimal value of $\alpha$ through techniques such as L-curve analysis \citep[e.g.][]{Hansen1992,Hansen1993} or cross-validation \citep[e.g.][]{Golub1979,Hammond2024}, we instead tune $\alpha$ such that the regularized model's mean squared error (MSE) matches that of the statistically-preferred eigenmapping fit. Unlike in the eigenmapping fit, where we restrict our model by the number of eigenmodes, this model includes \emph{all} possible eigenmodes and instead imposes a tunable prior on the model. For this paper, we generate a range of $\alpha$ values equally spaced in logarithm. We choose an initially broad range ($10^{-30}$ to $10^6$, with 61 points equally spaced in logarithm) to identify a lower limit where the power in higher degrees begins to dominate, and to identify the order of magnitude where the penalized model's MSE crosses that of the eigenmapping fit. We then generate a set of models more finely sampled in this narrower range (two orders of magnitude centered at the order where the MSE crosses the eigenmapping value), and select the penalty whose MSE is closest to the target. The complexity-penalized model is our preferred estimate of the emission map, and we compare it with our eigenmapping baseline for the fits to both artificial and real data in the rest of this work.

\section{Injection-Recovery Tests on Simulated Maps Varying Only in Latitude}
\label{sec:model_testing_in_latitude}
As preparation for testing our model framework on both simulated and real data, we examine how our model fits two simulated maps that vary only in latitude. By this construction, the companion has constant flux outside of eclipse. We do not repeat the transit fit as in Section \ref{sec:model_methods:orbital_fit}; the orbital geometry, host flux, and both host and companion radii are assumed to be known ahead of time as the maximum-likelihood values in Table \ref{table:orbital_fit_results}. We solve for the flux of the brown dwarf alone by subtracting the contribution of the white dwarf assuming the maximum-likelihood model for the WD emission. We set the noise level for the test directly from the per-point uncertainties of the wavelength-integrated data, without the $\times 1.4$ error inflation inferred from our orbital fit. This noise level corresponds to an eclipse-depth SNR of $\approx 6$ (see Section \ref{sec:WD1202:observations}), with all noise assumed to be Gaussian and un-correlated in time. This section considers two cases: a pure dipole \ref{sec:model_testing_in_latitude:dipole_test}, and an equatorially-concentrated band (hereafter the ``ring'' case, in Section \ref{sec:model_testing_in_latitude:equatorial_ring_test}).

\subsection{North-South Dipole Test: Method Retrieves Brightness Deviation in Occultation}\label{sec:model_testing_in_latitude:dipole_test}
A very simple case of latitudinal asymmetry in our harmonic framework is a $\ell=1$ dipole. This case is designed to produce zero astrophysical variation in the observed flux outside of eclipse. We parametrize our dipole strength with a latitudinal flux asymmetry metric
\begin{equation}\label{eq:dipole_strength}
    D \equiv \frac{F_\mathrm{N, obs} - F_\mathrm{S, obs}}{F_\mathrm{N, obs} + F_\mathrm{S, obs}}
\end{equation}
where the observed northern hemispherical flux is evaluated as
\begin{align}
    F_\mathrm{N, obs} &= \int_{-\pi/2}^{\pi/2} \int_{0}^{\pi/2} \sum\limits_{l, m} f_{lm} Y_{lm}\!\left(\theta, \phi\right) \, \sin^2\!\theta d\theta d\phi \\
    &\propto \int_{0}^{\pi/2} \left(f_{0, 0} Y_{0, 0} + f_{1, d} Y_{1, d}\right) \, \sin^2\!\theta d\theta \\
    &\propto \frac{\pi}{4} f_{0, 0} + \frac{1}{\sqrt{3}} f_{1, d}.
\end{align}
Here we use the normalization constants and real spherical harmonics used in \texttt{jaxoplanet}, where $Y_{0, 0} \equiv 1/\sqrt{4\pi}$. The choice of $d$ to denote the north--south $m$ value is intentional, as \texttt{jaxoplanet} puts the default polar axis for spherical harmonics along the line of sight to the observer ($d=-1$). If we assume the polar axis of the mapped object is in the plane of the sky, save for the orbital inclination, then $d=0$. For the southern hemisphere, there is a sign change
\begin{equation}\label{eq:dipole_southern_hemisphere_flux}
    F_\mathrm{S, obs} \propto \frac{\pi}{4} f_{0, 0} - \frac{1}{\sqrt{3}} f_{1, d}.
\end{equation}
To ensure there are non-negative fluxes at all observed latitudes, we restrict ourselves to $\left\lvert D \right\rvert < 4 / \!\left(3 \pi\right)$, which corresponds to $\left\lvert f_{1, d}/f_{0, 0} \right\rvert < 1 / \sqrt{3}$. For our dipole test, we arbitrarily set $D=0.4$. The resulting map changes the eclipse depth expected from a uniform map by $\approx 0.4$\%, as seen in the left panel of Figure \ref{fig:dipole_test_schematic_and_fit}.

\subsubsection{Eigenmapping the North-South Dipole: Detecting a Difference from the Eclipse of a Uniformly Bright Map}\label{sec:model_testing_in_latitude:dipole_test:eigenmapping}
\begin{figure*}
\begin{center}
\includegraphics[width=17cm]{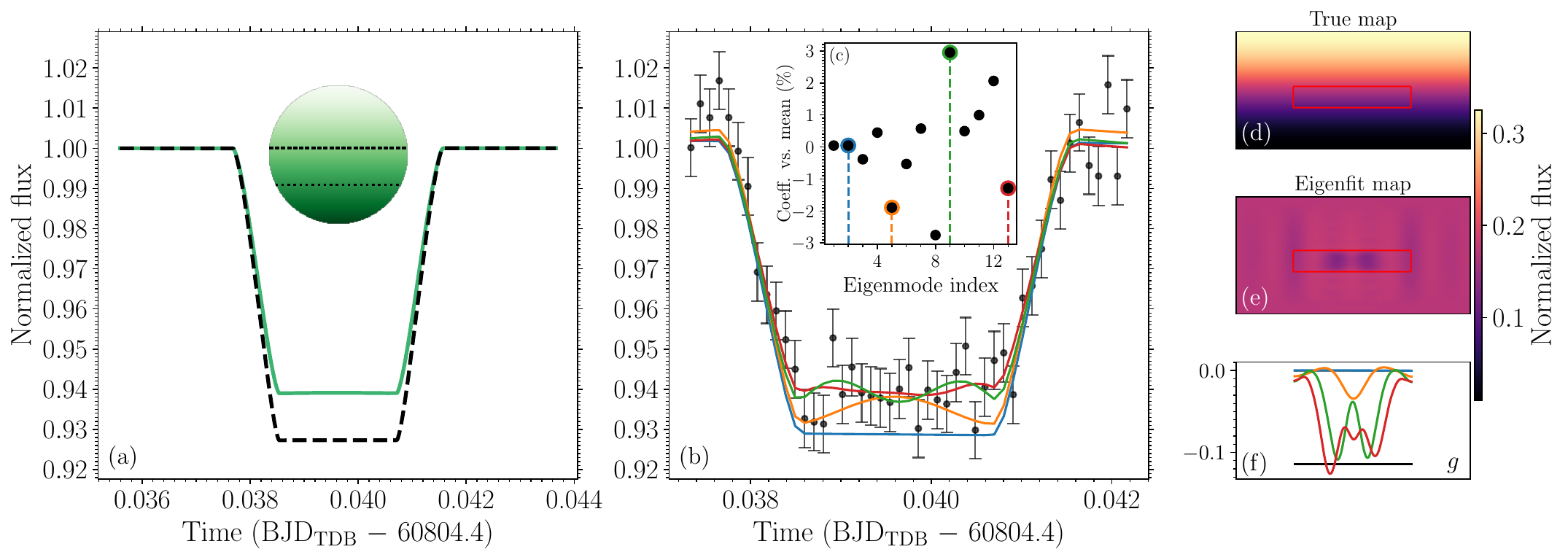}
\caption{\emph{(a):} A schematic showing the dipole test case, with the light curve during secondary eclipse of the true map (green), versus the light curve expected from a brown dwarf with uniform emission (black, dashed line). Each light curve is normalized to the $Y_{0,0}$ component of the map that produces it. Inset is a color map showing the projected map in the plane of the sky, with the lightness of the color scaled to increase with the flux. A pair of dotted black lines denote the latitude range of the path of the host during secondary eclipse, here computed for the $b \approx 0.64$ orbital geometry.  \emph{(b):} The results of the fit to the light curve in eclipse, with a selection of models that differ in the number of eigenmodes used. The statistically-preferred fit uses 9 eigenmodes and is shown in green. In the inset, we show the contribution of each mode relative to the mean brightness (constant term). We highlight a low-order fit in blue and a high-order fit in brown. In between, the highlighted colored modes correspond to where we gain structures that impart appreciable contribution to the flux in the occulted region. \emph{(d)--(e):} The input map and retrieved map, respectively, on the same color scale. The red box shows the boundary of the latitudes and longitudes that are ever occulted by the WD. \emph{(f):} The occultation asymmetry metric $g$, as defined in Equation \ref{eq:occultation_asymmetry_metric}, is plotted across the day-side longitudes in the upper row for a selection of model fits by number of eigenmodes. The true value of $g$ for the dipole map is shown as a black horizontal line.}
\label{fig:dipole_test_schematic_and_fit}
\end{center}
\end{figure*}

The results of our dipole map injection and recovery test are shown in the right panel of Figure \ref{fig:dipole_test_schematic_and_fit}. The best-fit emission map from the eigenmapping model returns a nearly uniform map except along the path of the occultation, where the flux is slightly dimmer. Since the WD transits across more of the dimmer hemisphere than the brighter hemisphere, the model ``sees'' that the occulted region is on average dimmer than what would be expected if the emission map were at a constant brightness tuned to the out-of-eclipse flux level. Figure \ref{fig:dipole_test_schematic_and_fit} shows how the fit quality improves as successive eigenmodes are added. We see the first substantial increase in fit quality at 5 eigenmodes, where the model gains access to a map with significant eclipse structure. The next substantial increase occurs at the inclusion of the 9th eigenmode which allows the model to change the eclipse depth with relatively little change to the phase variations. The fit that is statistically preferred for our test by a BIC test uses 9 eigenmodes.

These results suggest the simplest hypothesis test for our eclipse mapping model is one that asks whether the average flux across the occulted region deviates from what one expects from fitting the phase curve alone. Here recall we have purposefully isolated a structure whose structure can only be probed with an eclipse, since the dipole map has zero phase variation. We can define a metric that accounts for \emph{occultation} asymmetry rather than north-south hemisphere asymmetry. Call the expected observed flux for a uniform map within the occulted region $F_\mathrm{in, unif.}$ and the expected observed flux outside the region $F_\mathrm{out, unif.}$. The actual observed fluxes are denoted as $F_\mathrm{in}$ and $F_\mathrm{out}$, respectively. Then
\begin{equation}\label{eq:occultation_asymmetry_metric}
    g \equiv \frac{\frac{F_\mathrm{in}}{F_\mathrm{in, unif.}}-\frac{F_\mathrm{out}}{F_\mathrm{out, unif.}}}{\frac{F_\mathrm{in}}{F_\mathrm{in, unif.}}+\frac{F_\mathrm{out}}{F_\mathrm{out, unif.}}}
\end{equation}
with $g \rightarrow 0$ denoting an observed flux perfectly consistent with the expectation from a uniform brightness, $g \rightarrow +1$ implying \emph{all} the observed flux is concentrated entirely within the occulted region, and $g \rightarrow -1$ implying none of the observed flux is contained in the occulted region. For our input map, $g \approx -0.11$ across all longitudes. Our model should be able to recover a non-zero $g$ across the day side, and ideally recover an accurate estimate. We show the value of this metric as a function of day-side longitude and for selected numbers of included eigenmodes in Figure \ref{fig:dipole_test_schematic_and_fit}. The BIC-preferred fit (9 eigenmodes) detects a significantly non-zero $g$ on the day side, but under-estimates the true value. This under-estimation persists when a higher number of eigenmodes is used, as well as when the simulated SNR is increased (not shown). Since the eigenmapping model is designed to use map structures that are ``efficient'' in the occulted region, it needs to inject less contrast to reproduce the dipole's eclipse depth than actually exists in the dipole map itself.

\subsubsection{The Dipole Test with a Complexity-Penalized Model}\label{sec:model_testing_in_latitude:dipole_test:complexity_penalization}
\begin{figure*}
\begin{center}
\includegraphics[width=17cm]{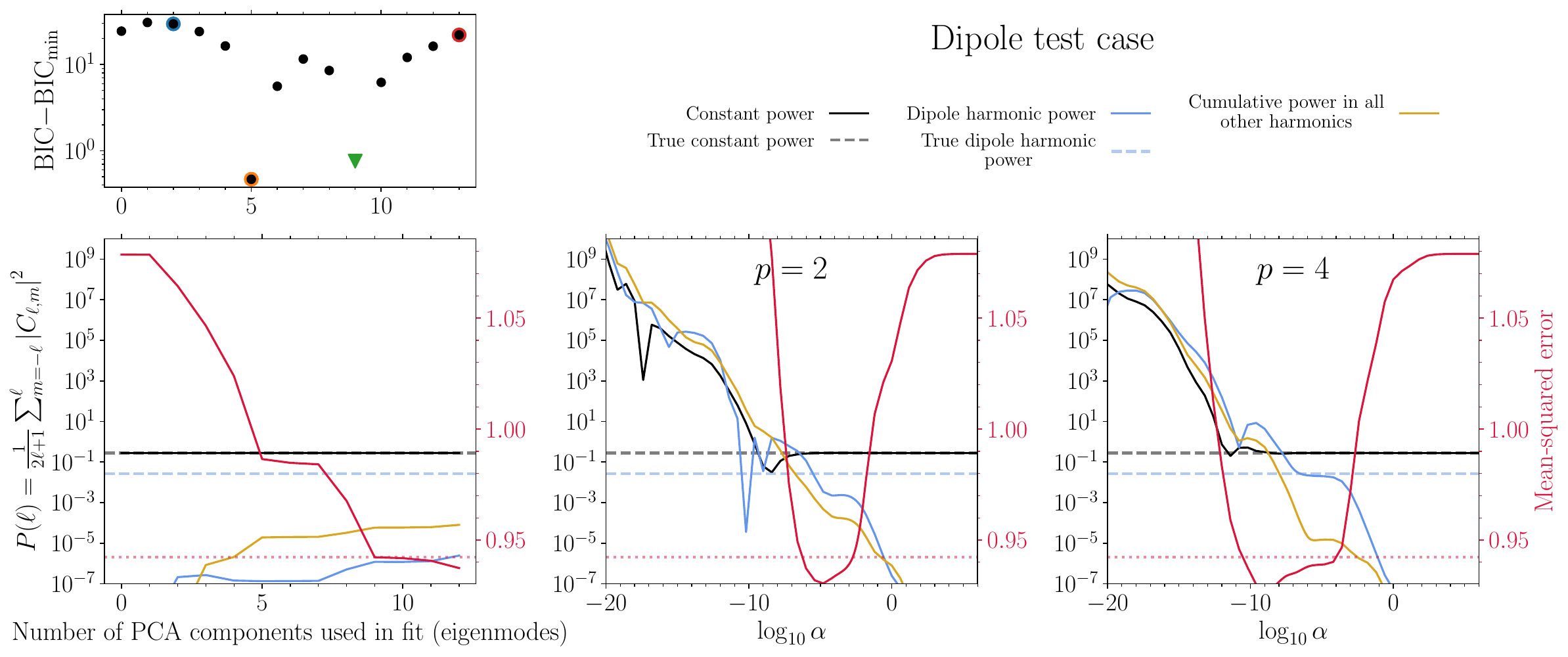}
\caption{A comparison of the components of the harmonic power spectrum at the constant (i.e.~$Y_{0, 0}$) mode (in black), first degree (in blue), and the cumulative power at all subsequent degrees $\ell \geq 2$ (in gold) for the fits to the dipole map both with the eigenbasis (lower left panel), which is tuned by the number of included eigenmodes; and the complexity-penalized model (lower right panel). The upper left panel shows the statistical metric, the Bayesian Information Criterion (BIC, as described in Section \ref{sec:model_methods:model_selection}), as a function of the number of eigenmodes. Note that the colors of selected eigenmodes correspond to those shown in Figure \ref{fig:dipole_test_schematic_and_fit}. For the dipole test, we find a minimum (optimal) value at 9 eigenmodes, as indicated by the triangle icon pointing to the minimum at zero. The penalized fit is then selected at the $\alpha$ whose MSE matches that of the BIC-preferred eigenmapping fit, the magnitude of the penalty term used in the complexity-penalized model. $p$ is the exponent of the penalty that controls the prior on how the harmonic power spectrum falls off with increasing degree $\ell$. The initial approach is to fix $p$ at a value of 2, which yields second-order Tikhonov regularization. To show that the model can converge to the true map in this simple example, we show a case of a stricter penalty exponent ($p=4$).}
\label{fig:dipole_penalized_fit_power_spectra}
\end{center}
\end{figure*}

\begin{figure}
\begin{center}
\begin{tabular}{cc}
\includegraphics[height=6cm]{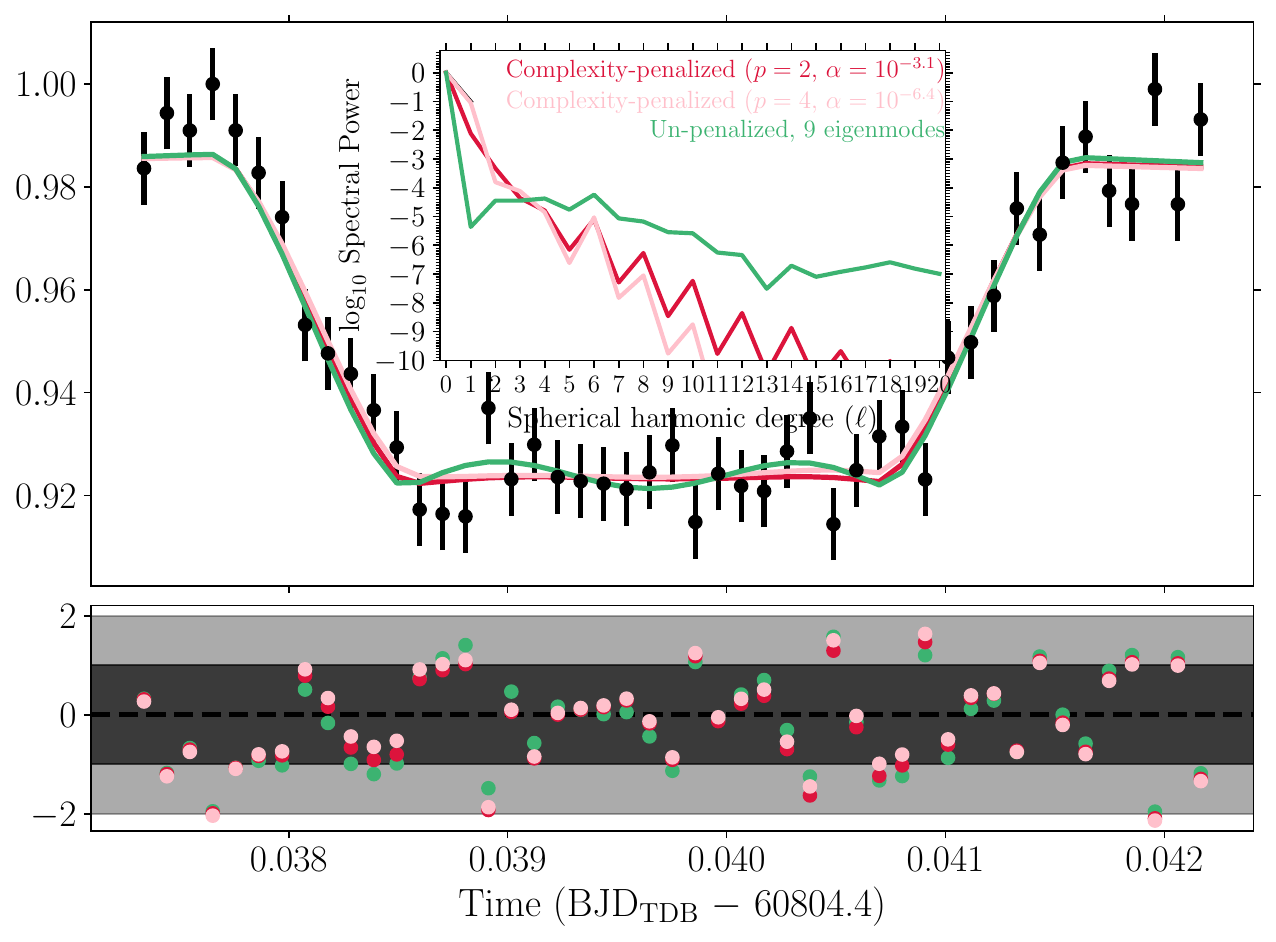} &
\includegraphics[height=6cm]{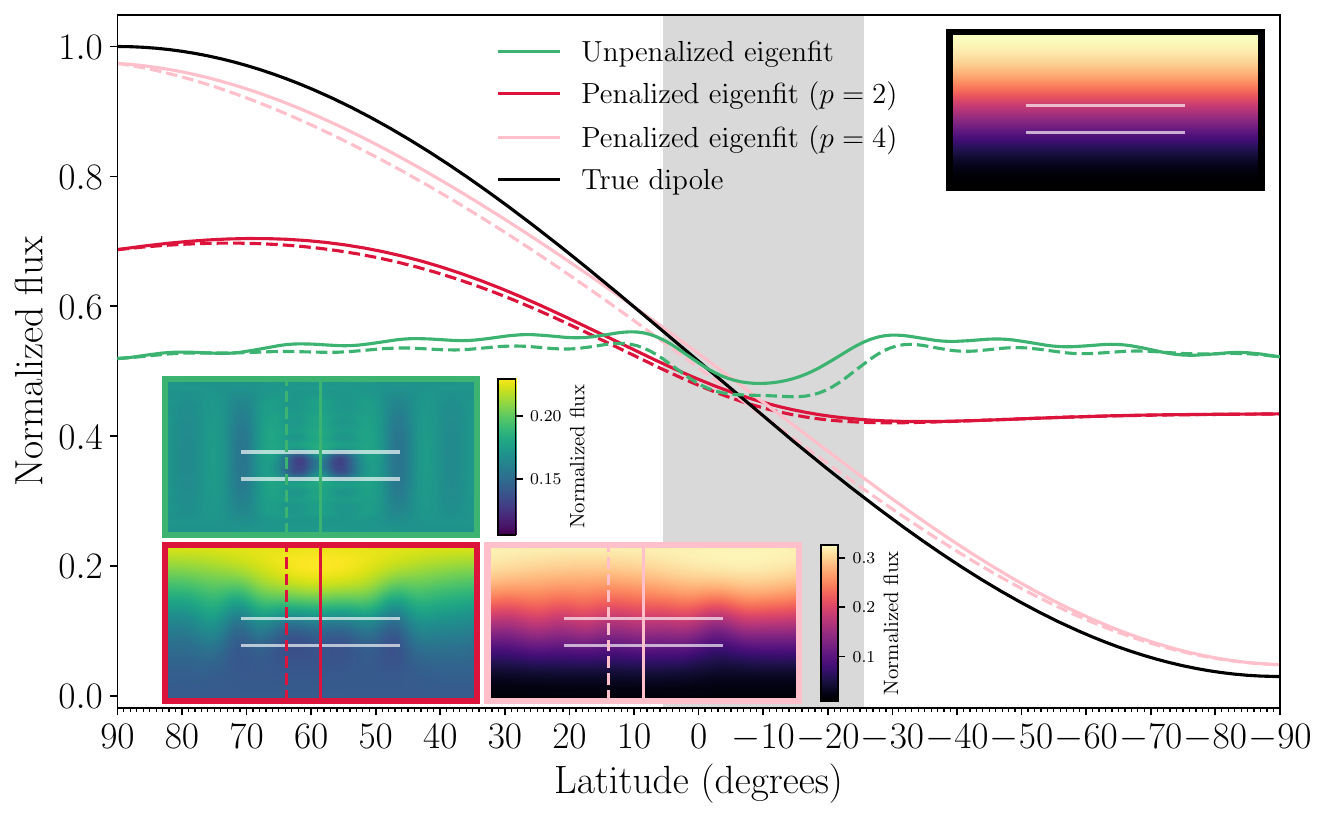} \\
\end{tabular}
\caption{A comparison of the best fit to the dipole map with the eigenbasis (green), versus a fit where the harmonic power is penalized as a function of degree (red for $p=2$ and pink for $p=4$, as detailed in Section \ref{sec:model_methods:complexity_penalization}). We select the complexity penalty such that the penalized model's MSE matches that of the eigenbasis fit; the eigenbasis fit uses 9 eigenmodes as determined by the model selection (BIC comparison). \emph{Left:} the models versus data within secondary eclipse, with scaled residuals shown in the lower panel. Here the eclipse is normalized by subtracting the contribution from the WD (as modeled in the initial orbital fit), and dividing by the maximum of the data. In the inset plot, we show the spherical harmonic power spectra for the true map and the two model fits. \emph{Right}: meridional slices of the emission fluxes for our fits, as well as the true flux, along with their corresponding maps. The parallel light grey lines show the boundaries of where the map is occulted. The solid curves represent a slice at the sub-stellar (zero) longitude and the dashed lines represent slices $40^\circ$ away from the sub-stellar longitude. The two color scales represent the same quantities, with the ``viridis'' scale using a narrower dynamic range to more clearly show the features of the un-penalized and penalized maps.}
\label{fig:dipole_penalized_versus_unpenalized_fits}
\end{center}
\end{figure}

We now apply the complexity penalty model of Section \ref{sec:model_methods:complexity_penalization} to our dipole test case. The dipole map has power in exactly two components, $Y_{0, 0}$ and $Y_{1, -1}$, which means its power spectrum is artificially truncated. We expect for this contrived case that a sufficiently (but perhaps unrealistically) high amount of penalization can recover the true dipole. Figure \ref{fig:dipole_penalized_fit_power_spectra} shows the spectral power of the fits, defined as $P\!\left(\ell\right) = \frac{1}{2 \ell + 1}\sum_{m=-\ell}^{\ell} \left|C_{\ell, m}\right|^2$, for the constant and dipole degrees ($\ell = 0$ and 1), along with the cumulative power in all remaining degrees, as a function of the two penalty parameters $\alpha$ and $p$. We start with the default choice of $p=2$. We include values of $\alpha$ down to $\sim 10^{-20}$ to show the behavior of the model when it is allowed to overfit: the model is free to choose very high-degree terms at very high powers, generating a map that goes to extreme positive and negative fluxes. Values of $\alpha$ that allow the power in non-zero degrees to be much higher than the constant power are unlikely to represent a physically plausible fit and therefore set a heuristic lower bound. As we tune $\alpha$ higher, the mean-squared error (MSE) steadily rises and the power in all degrees falls. By $\alpha \sim 10^{-5}$, the model's mean brightness has converged to the true value, and remains converged through the strongest penalties probed. The equal-MSE selection occurs at $\alpha \sim 8.5\times10^{-4}$ which is therefore our choice for the penalized fit. While the dipole coefficient is not retrieved quite accurately, the day side of the retrieved map shows a very similar north-to-south flux contrast (Figure \ref{fig:dipole_penalized_versus_unpenalized_fits}). Like the eigenmapping fit, the retrieved north-south gradient is considerably smaller than that of the true map. There is a small but non-negligible power at $\ell > 1$ degrees, which is a consequence of the relative permissiveness of the power spectrum penalty relative to the very sharp cut-off of the simulated map's power spectrum.

We show the model is capable of converging on the true value for a simplified case by penalizing higher-degree modes more severely. We can do this by increasing the penalty exponent $p$ beyond 2; Figure \ref{fig:dipole_penalized_fit_power_spectra} shows the case for $p=4$. At $\alpha \sim 10^{-7}$ to $10^{-6}$, we see both the constant and dipole powers level to nearly the true values, with the MSE-matched fit at $\alpha \sim 4\times10^{-7}$. We note that the specific choice of exponent $p>2$ is an arbitrary choice that is designed to show that there exists a solution with a sufficient penalization to converge on a solution that approaches the dipole when choosing $\alpha$ by matching the un-penalized MSE. It is unclear whether increasing the penalty exponent beyond $p=2$ would be useful to fit physically-realistic maps in the absence of a physically-motivated prior.

\subsection{Equatorial Ring Test: Flux Concentration Affects Inferred Brightness}\label{sec:model_testing_in_latitude:equatorial_ring_test}
We now test whether our model accurately estimates the occultation metric for a simulated map where flux is concentrated toward the BD's equator. As with the dipole, there is zero phase variation outside eclipse. We modify the \texttt{ring\_y} function in \texttt{jaxoplanet}, which was under construction as of the writing of this work, to create the map. This function is adapted from the sigmoid spot profile originally used in \texttt{starry} \citep{Luger2019}, here to set the brightness profile in latitude relative to the brown dwarf's equator. The width scale of the ring is set to 10 degrees as a rough approximation of the scale of a jet on a rapidly-rotating irradiated brown dwarf. The brightness of the injected ring is arbitrarily set to drop to 0.4 times the peak equatorial flux by $\pm 20^\circ$ latitude. As with the north-south dipole test, the light curves are simulated with the un-inflated per-point noise level of the real data (eclipse-depth SNR $\approx 6$).

\begin{figure*}
\begin{center}
\includegraphics[width=17cm]{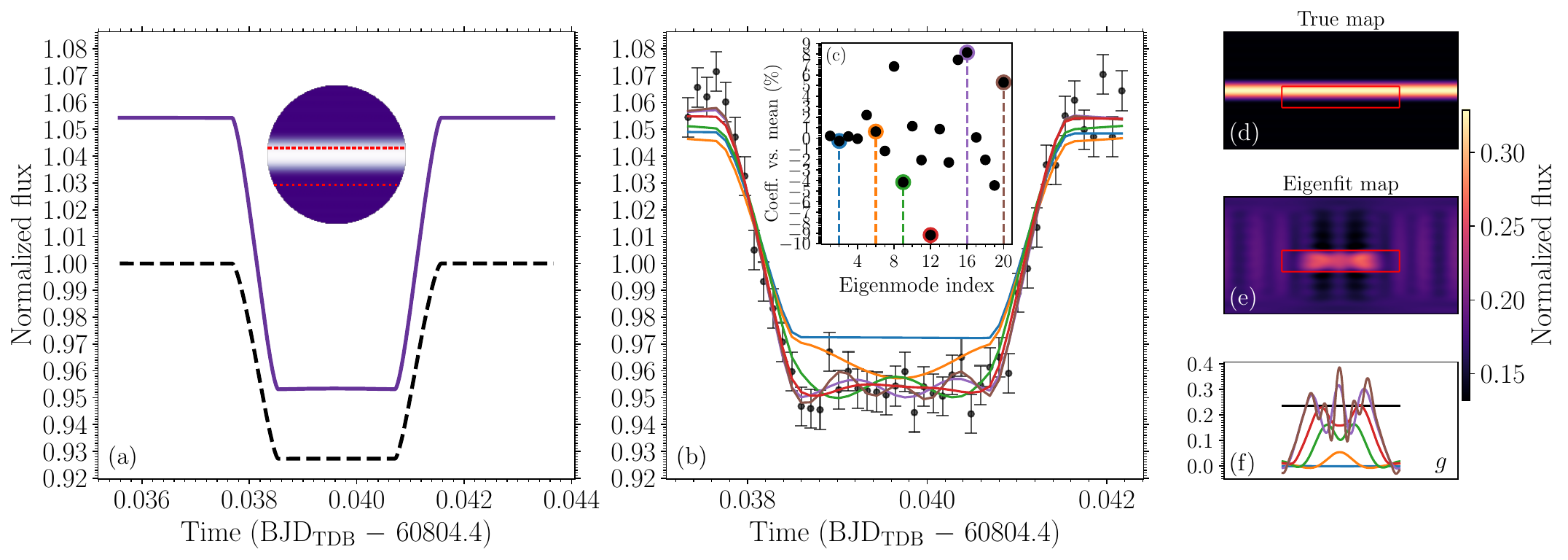}
\caption{\emph{(a):} A schematic showing the equatorial ring test case, with the light curve during secondary eclipse of the true map (purple), versus the light curve expected from a brown dwarf with uniform emission (black, dashed line). Each light curve is normalized to the $Y_{0,0}$ component of the map that produces it. Inset is a color map showing the projected map in the plane of the sky, with the lightness of the color scaled to increase with the flux. A pair of dotted red lines denote the latitude range of the path of the host during secondary eclipse, here computed for the $b \approx 0.64$ orbital geometry. \emph{(b):} The results of the fit to the light curve in eclipse, with a selection of models that differ in the number of eigenmodes used. The statistically-preferred fit uses 12 eigenmodes and is shown in red. In the inset, we show the contribution of each mode relative to the mean brightness (constant term). We highlight a low-order fit in blue and a high-order fit in brown. In between, the highlighted colored modes correspond to where there we gain structures that impart appreciable contribution to the flux in the occulted region. \emph{(d)--(e):} The input map and retrieved map, respectively, on the same color scale. The red box shows the boundary of the latitudes and longitudes that are ever occulted by the WD. \emph{(f):} The occultation asymmetry metric $g$, as defined in Equation \ref{eq:occultation_asymmetry_metric}, is plotted across the day-side longitudes in the upper row for a selection of model fits by number of eigenmodes. The true value of $g$ for the ring map is shown as a black horizontal line.}
\label{fig:ring_test_schematic}
\end{center}
\end{figure*}

We immediately notice that the baseline flux of the light curve outside eclipse is $\approx 5.5$\% higher (central panel of Figure \ref{fig:ring_test_schematic}) than would be expected from a uniform map with the same mean global brightness. This is a projection effect, with concentration of flux at latitudes more aligned with our nearly edge-on line of sight. This projection effect is not unique to our equatorial ring since it would also be present, for example, on the day side of a map with a day-night contrast. The key difference for our equatorial ring map is that, outside of eclipse, the effect is completely hidden in the harmonic ``null space'': with just the rotational phase curve, we cannot distinguish our equatorial ring map from a uniform map with a $\approx 5.5$\% higher mean brightness. We note that this effect is unlikely to be as severe as is seen in this artificial test case, since a more realistic companion map would have day-night phase variations of much higher amplitude than the contribution in emission from an equatorial jet. It will nevertheless be present in any map whose latitudinal brightness profile shows a net concentration toward the equator.\footnote{The net concentration should technically be toward latitudes more aligned with our line of sight. However, for a companion with assumed zero obliquity and a nearly edge-on view, these are very similar conditions.}

\subsubsection{The Eigenmapping Method Fits the Eclipse but Over-Estimates the Mean Brightness}\label{sec:model_testing_in_latitude:equatorial_ring_test:eigenmapping}

Our eigenmapping fit results for the equatorial ring case are shown in the central and right columns of Figure \ref{fig:ring_test_schematic}. Here the retrieved map identifies the flux of the ring along the eclipse path, and we see a similar punctuated increase in the fit quality as the number of basis light curves is increased and the model reaches eigenmodes that can refine the fit spatially. At our simulated SNR we find 12 eigenmodes is statistically preferred. As in the dipole case, the inferred occultation asymmetry metric does not reach the true value of $g \approx 0.24$ for the statistically-preferred model, though fits with more eigenmodes do approach and oscillate around the true value.

Our preferred eigenmapping model over-estimates the mean brightness of the true map by $\approx 4.4$\%. This stems from a few causes: (i) the concentration of flux from our model somewhat off the equator due to the non-zero impact parameter; (ii) the ``effective beam width'' of the WD's occultation is a bit broader in latitude than the nominal width of the jet, though only slightly; and (iii) because the model is insensitive to the equatorial concentration of flux on the night side. Repeating this model test with $b_\mathrm{MLE} = 0.015$ (not shown) results in an over-prediction of $\approx 3.5$\%. The spectral power trends also show that more power is retrieved at the first harmonic degree than is present, and the models under-estimate the remaining combined power across all higher degrees. This illustrates that the eigenmapping model is producing a \emph{localized} fit to the occultation; the model finds a valid solution with a higher mean brightness, as increasing the zeroth eigenmode coefficient and down-weighting higher-order modes is the most efficient way to reproduce the light curve with our eigenbasis.

\subsubsection{The Equatorial Ring Test with a Complexity-Penalized Model}\label{sec:model_testing_in_latitude:equatorial_ring_test:complexity_penalization}
\begin{figure}
\begin{center}
\includegraphics[width=17cm]{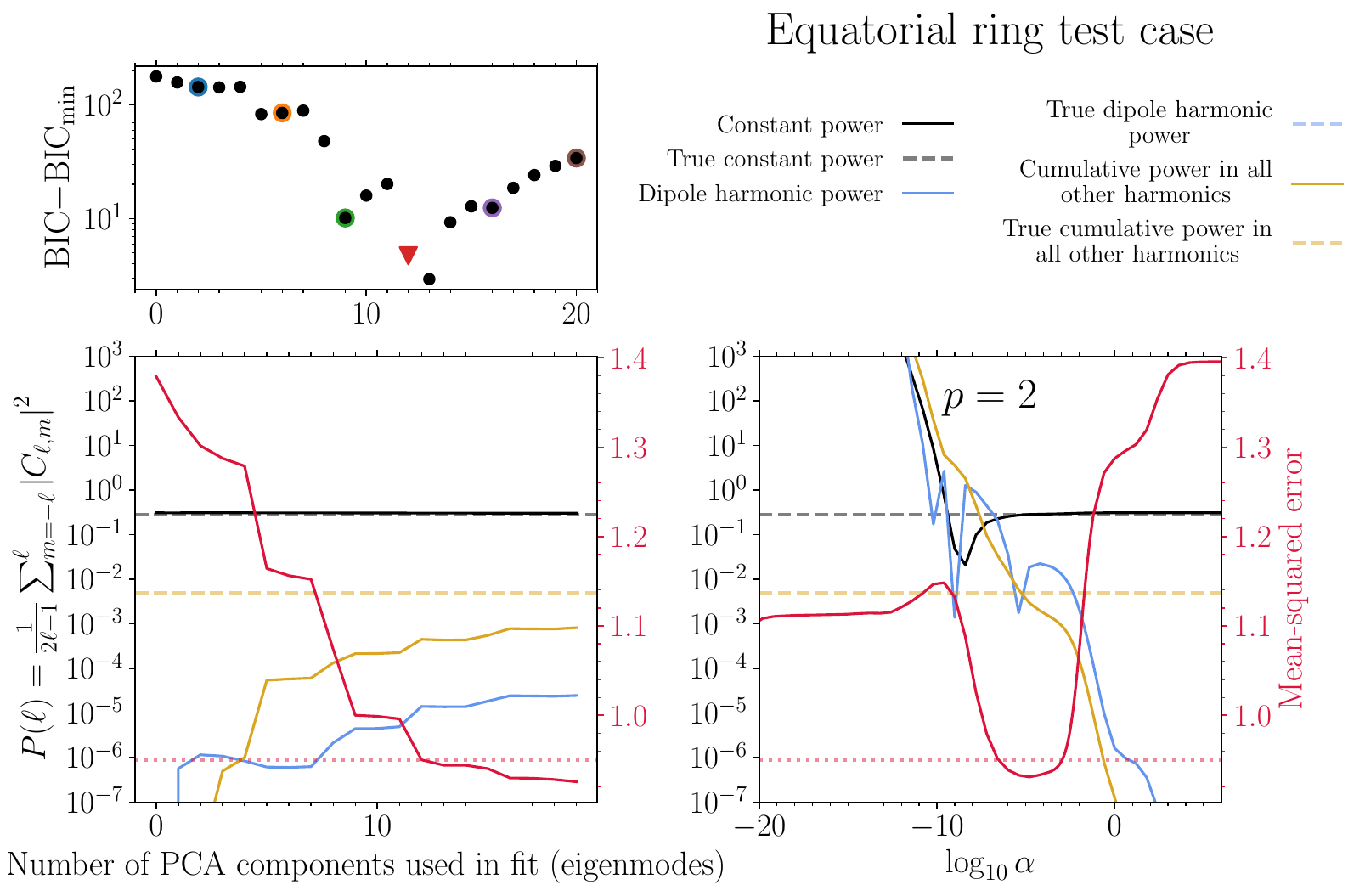}
\caption{A comparison of the components of the harmonic power spectrum at the constant (i.e.~$Y_{0, 0}$) mode (in black), first degree (in blue), and the cumulative power at all subsequent degrees $\ell \geq 2$ (in gold) for the fits to the equatorial ring map both with the eigenbasis (lower left panel), which is tuned by the number of included eigenmodes; and the complexity-penalized model (lower right panel). The upper left panel shows the statistical metric, the Bayesian Information Criterion (BIC, as described in \ref{sec:model_methods:model_selection}), as a function of the number of eigenmodes. Note that the colors of selected eigenmodes correspond to those shown in Figure \ref{fig:dipole_test_schematic_and_fit}. For the ring test, we find a minimum (optimal) value at 12 eigenmodes, as indicated by the triangle icon pointing to the minimum at zero. The penalized fit is then selected at the $\alpha$ whose MSE matches that of the BIC-preferred eigenmapping fit, the magnitude of the penalty term used in the complexity-penalized model. $p$ is the exponent of the penalty that controls the prior on how the harmonic power spectrum falls off with increasing degree $\ell$.}
\label{fig:ring_penalized_fit_power_spectra}
\end{center}
\end{figure}

\begin{figure}
\begin{center}
\begin{tabular}{cc}
\includegraphics[height=6cm]{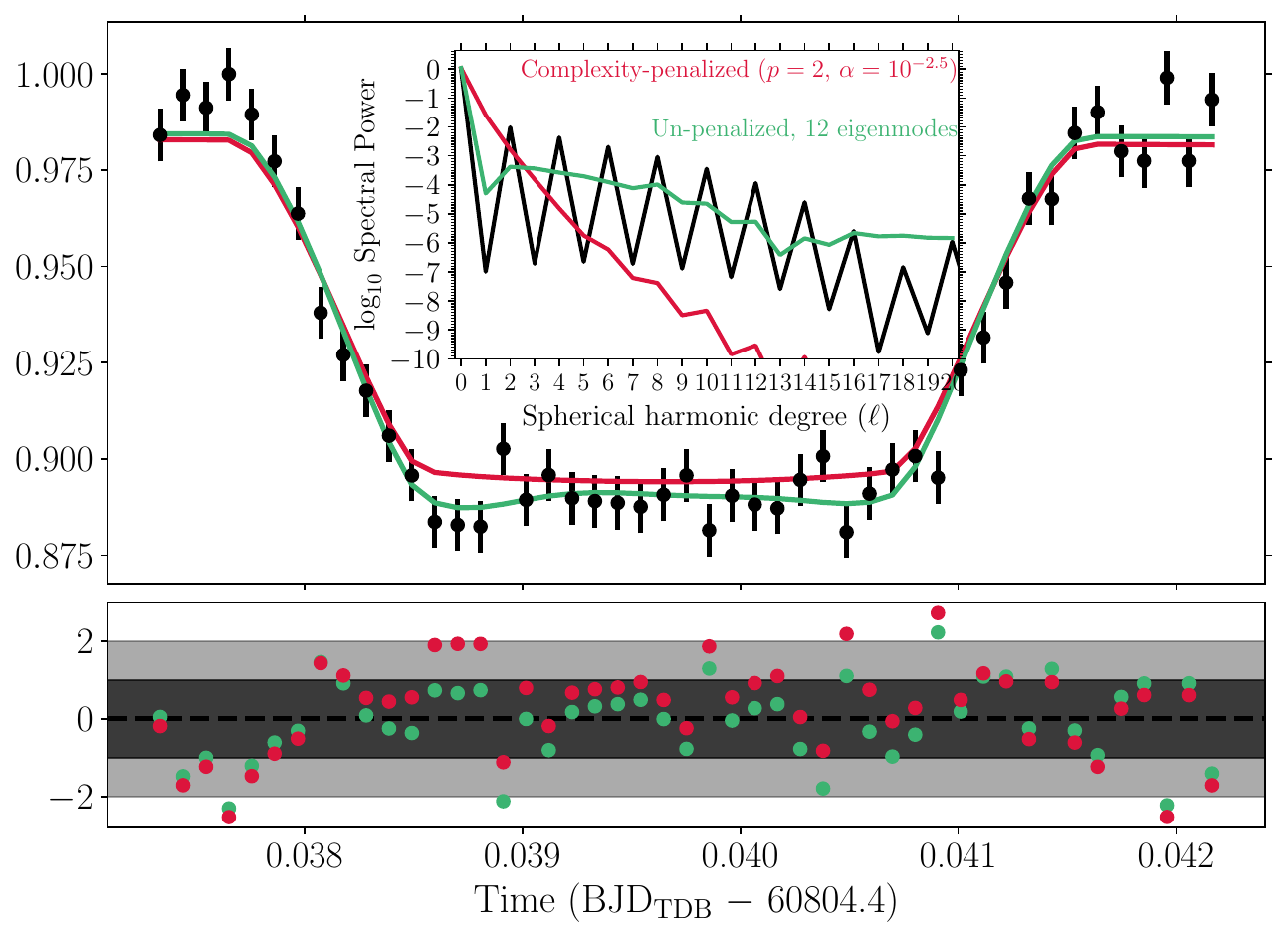} &
\includegraphics[height=6cm]{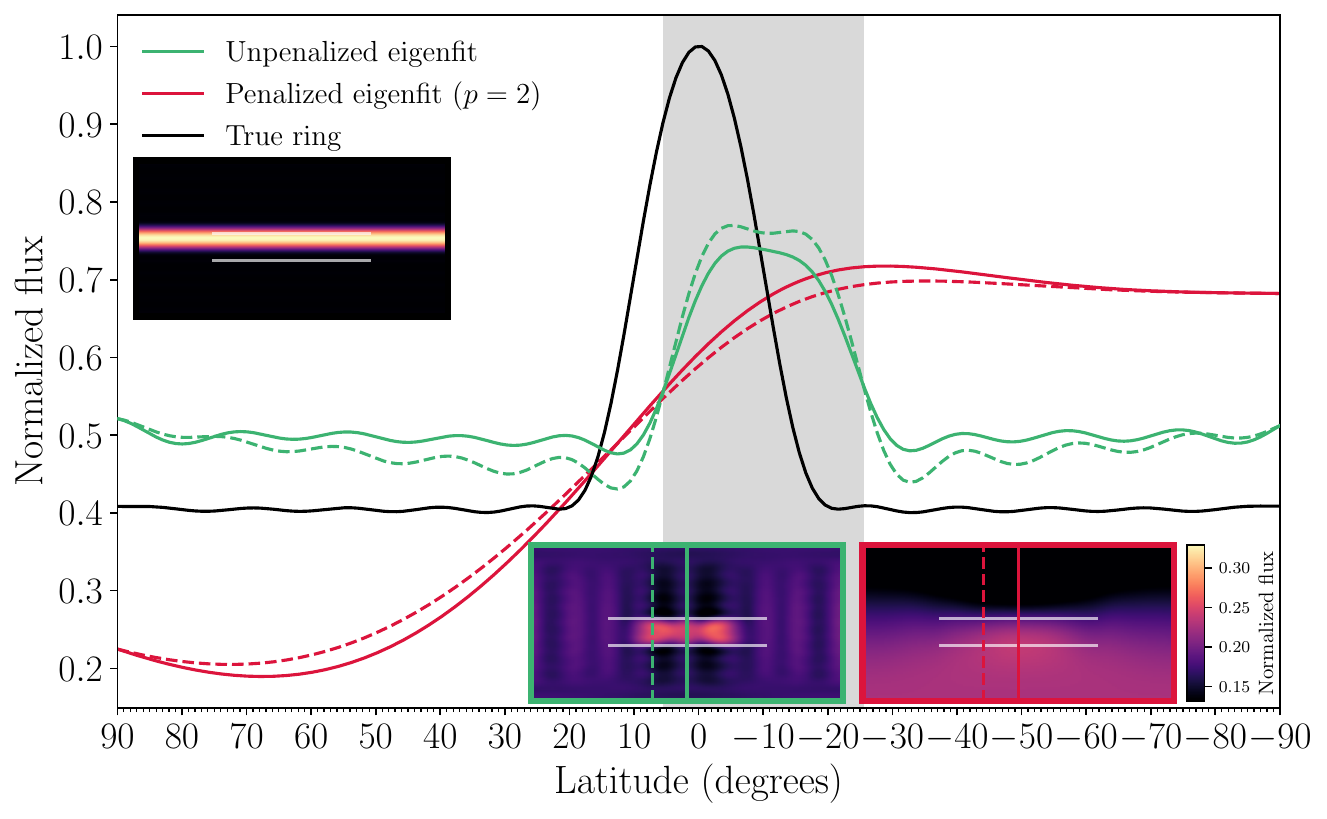} \\
\end{tabular}
\caption{A comparison of the best fit to the equatorial ring map with the eigenbasis (green), versus a fit where the harmonic power is penalized as a function of degree (red, as detailed in Section \ref{sec:model_methods:complexity_penalization}). We select the complexity penalty such that the penalized model's MSE matches that of the eigenbasis fit; the eigenbasis fit uses 12 eigenmodes as determined by the model selection (BIC comparison). \emph{Left:} the models versus data within secondary eclipse, with scaled residuals shown in the lower panel. Here the eclipse is normalized by subtracting the contribution from the WD (as modeled in the initial orbital fit), and dividing by the maximum of the data. In the inset plot, we show the spherical harmonic power spectra for the true map and the two model fits. \emph{Right}: meridional slices of the emission fluxes for our fits, as well as the true flux, along with their corresponding maps. The parallel light grey lines show the boundaries of where the map is occulted. The solid curves represent a slice at the sub-stellar (zero) longitude and the dashed lines represent slices $40^\circ$ away from the sub-stellar longitude. These values are chosen approximately to dissect the regions in both the un-penalized and penalized maps where the flux is preferentially concentrated.}
\label{fig:ring_penalized_versus_unpenalized_fits}
\end{center}
\end{figure}

In addition to the even-heavy harmonic power spectrum of the equatorial ring, the functional form of our ring model --- a sigmoid profile --- does not follow a power law trend in its power spectrum, but instead an exponential. Therefore, we should expect that, based on the noise floor of our data, we may find there is no ``true'' value of $p$ that fully optimizes our fit to our artificial map. Since we expect a ring-like structure to be just one component of a map, we do not expect an exponential drop-off in general, and it is instructive to examine how our complexity-penalized model reproduces such a feature. We start with our default of $p=2$ and select the penalty whose MSE matches that of the eigenmapping fit; our model selection criterion prefers to include 12 eigenmodes. This yields a favored value of $\alpha \sim 3\times10^{-3}$ (Figure \ref{fig:ring_penalized_fit_power_spectra}). There are no solutions in our complexity-penalized framework that recover the ring shape accurately. The retrieved map resembles a north-south dipole with a slight bias toward preserving flux along the occultation path on the day side (Figure \ref{fig:ring_penalized_versus_unpenalized_fits}).

Figure \ref{fig:ring_penalized_versus_unpenalized_fits} shows both the un-penalized and penalized light curve fits in eclipse (left panel) and the meridional (fixed longitude) slices of flux for the true map and both model fits (right panel). The penalized model adopts a flatter profile for the eclipse, fitting fewer points as closely as the un-penalized fit but making up for the difference by reaching closer to the higher-flux data points. This has no immediate physical interpretation, and is simply a consequence of our choice to match the mean-squared error of the un-penalized fit while minimizing harmonic complexity. There are two slight dips in the un-penalized light curve at the beginning and end of the portion of the eclipse where the WD's disk maximally overlaps with that of the BD. One can see this in the fit to the un-penalized eigenbasis, where it concentrates its flux within the occultation path. There are two bright lobes that peak at $\approx 40^\circ$ on either side of the sub-stellar longitude. This is a result of the model selection which finds a low-order polynomial shape is sufficient to model the data within eclipse. The amplitude of the flux change in the un-penalized map is slightly less than half that of the true ring; the flux stays about 25\% higher outside of the occultation latitudes. The complexity-penalized fit shows features broader in angular scale, with significant power allowed to bleed outside the occultation path while maintaining the fit quality in the light curve. Were one to imagine scanning the flux across latitudes, with the ring as a sort of impulse, the complexity-penalized profile resembles a cumulative response to the impulse. Its latitudinal profile shows the simplified gradient that resembles a north-south dipole, with the slight shift toward northerly latitudes required to inject flux along the occultation.

\section{Injection-Recovery Tests on the Predicted Emission of a WD-Irradiated BD Atmosphere}
\label{sec:GCM_test_results}

\begin{figure*}
\begin{center}
\includegraphics[width=17cm]{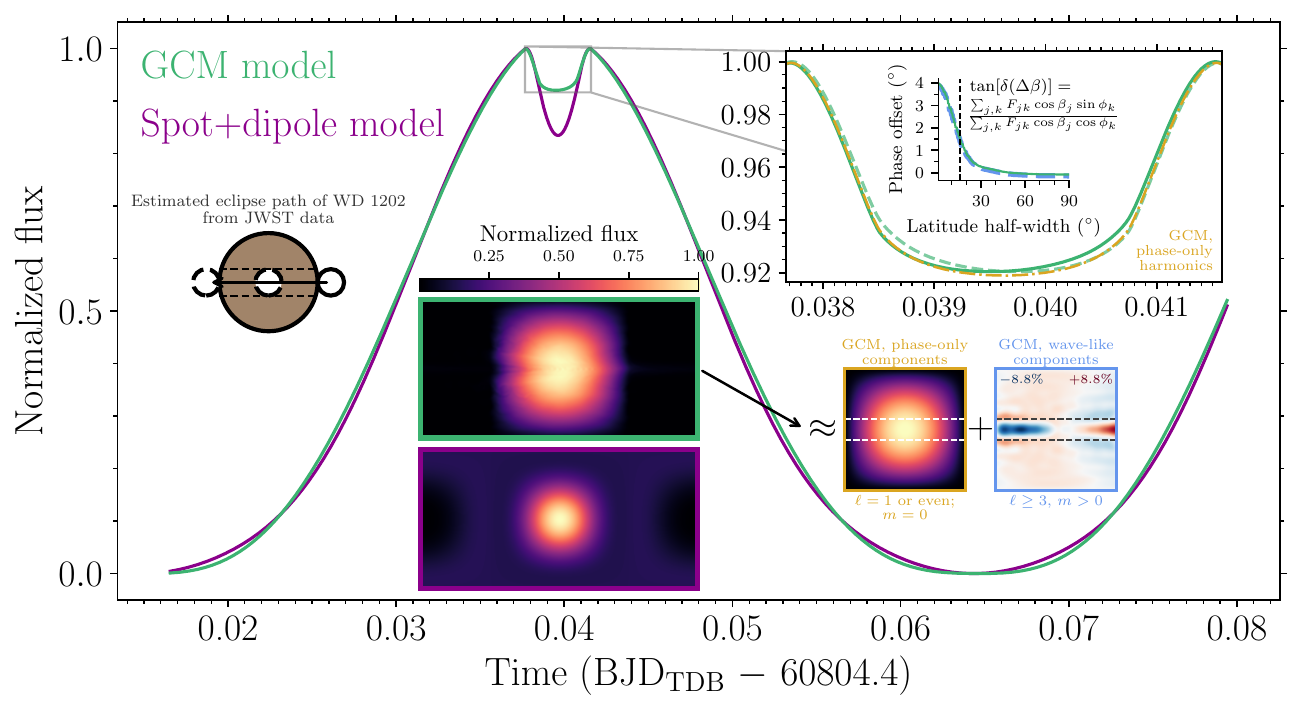}
\caption{\emph{Left:} reproduced light curves from the models of EPIC 2122 B in \citet{Zhou2022a} best-fit ``spot and dipole'' model (Section 3.4 in their paper; here reproduced in purple) and a scaled light curve from a GCM (green). We take the spherical harmonic transforms of each of these fits, place them on a simulated companion with the same eclipsing orbital geometry as WD 1202 b at the best-fit impact parameter $b\approx0.015$ (see the diagram to the left of the light curves), and show that the models predict noticeably different shapes in their eclipse light curves. Here each curve is normalized to its maximum flux. \emph{Inset, below eclipse:} the input maps, with equivalent color scales normalized to the maximum flux. The hotspot was constrained to $\sim 15^\circ$ in the original \citet{Zhou2022a} analysis, which reproduces the shape of the light curve's phase-only variations. However, the GCM-fit light curve produces a much shallower, flatter eclipse than that of the spot and dipole model. \emph{Inset, upper right:} a zoom-in on just the GCM light curve in eclipse, with the reflection of the light curve across the time of mid-eclipse as a dashed-line curve. The slight asymmetry comes from the longitudinal displacement of flux near the equator from the super-rotating jet. The gold dot-dashed curve is the eclipse of the phase-only harmonics alone, as defined in Section \ref{sec:GCM_test_results}. \emph{Inset, within eclipse:} the effective phase offset (Equation \ref{eq:effective_phase_offset}) of the GCM map (solid green line) and of its phase-only plus wave-like components (dashed blue line, also defined in Section \ref{sec:GCM_test_results}), with the half-width of the occulted band as a vertical dashed line. The effective phase offset peaks at $\approx 4^\circ$ at the equator, but falls to zero when integrating the flux across all latitudes which would render the feature invisible if this were a fully-occulted companion such as a hot Jupiter. \emph{Inset, lower right:} the day sides of the phase-only and wave-like components, with the occulted band dashed. The wave-like component map is plotted on a divergent color scale with a dynamic range of $\pm 8.8$\% of the peak day-side brightness (the top of the main color bar), with redder fluxes being higher flux; see also Figure \ref{fig:GCM_penalized_versus_unpenalized_fits}. These two harmonic sub-sets of the true map comprise all except $\approx 0.01\%$ of the non-constant power; $\approx 0.05\%$ is contained in the wave-like harmonics and the remaining $\approx 99.94\%$ is contained in the phase-only harmonics).}
\label{fig:GCM_curves_and_maps}
\end{center}
\end{figure*}

The latitudinal injection-recovery tests of Section \ref{sec:model_testing_in_latitude} demonstrate that our eigenmapping framework can retrieve isolated latitudinal structure from simulated eclipse data. A more realistic emission map will contain both longitudinal and latitudinal structure --- the former driven principally by the day-night contrast --- and the challenge is to determine the presence and structure of features in the light curve that can only come from latitudinal variations. For this next test, we start with a light curve generated from the emission map predicted from an existing atmospheric model of EPIC 2122 B \citep{Zhou2022a}. With a day-side temperature of order $\sim 4000$ K and an orbital period of 68.21 minutes \citep{Casewell2018a}, we expect the atmospheric structures of EPIC 2122 B to be comparable to those of WD 1202 B. \citet{Zhou2022a} originally model the phase variations with a simple empirical model consisting of a sigmoid spot profile on top of an $\ell=1$ dipole whose maximum is fixed to the sub-stellar point. We hereafter refer to this as the ``spot and dipole'' model. The spot and dipole model has four free parameters (mean brightness, dipole coefficient, hotspot intensity, and hotspot angular radius) that are fit directly to the data. \citet{Zhou2022a} then produce a second model light curve using the predicted emission from a GCM adapted from \citet{tanAtmosphericCirculationUltrahot2019} with parameters fixed to values drawn from models of similar irradiation temperatures as EPIC 2122 B in \citet{Lothringer2020}. The best-fit phase curve from the GCM is the fixed-parameter model whose top-of-atmosphere emission phase curve is scaled to the maximum and amplitude of the data.

Both model phase curves return similar fit qualities for EPIC 2122, which is a non-eclipsing system. However, the model predictions differ considerably when the system geometry is transiting (Figure \ref{fig:GCM_curves_and_maps}). We adopt the best-fit orbital geometry of WD 1202 from Section \ref{sec:model_methods:orbital_fit}. The $\sim 15^\circ$ hotspot radius of the original \citet{Zhou2022a} fit reproduces the ratio of the second-order to first-order terms in the Fourier series of the phase curve, but produces a sharp dip in the light curve in secondary eclipse when compared with the light curve of the GCM. There is also a slight asymmetry in the shape of the GCM eclipse where the first half (post-ingress) is deeper on average than the second half (pre-egress); we will refer to this as an ``early--late'' asymmetry. This comes from the eastward advection of flux on the day side near the equator which can be seen in the broad-band emission map. The large overlap between the latitudes the WD occults during secondary eclipse and the latitudes of the equatorial jet means that the secondary eclipse's early--late asymmetry is a magnified version of the narrow-latitude phase offset.

\subsection{Defining an ``Effective'' Phase Offset}\label{sec:GCM_test_results:effective_phase_offset}
Building on this idea, we define an ``effective'' phase offset that is analogous to the hemisphere-wide longitude shifts in the peak brightness seen in hot Jupiters. This is the mean flux-weighted longitude within a latitude window that extends $\Delta \beta$ in latitude in either direction from the equator. We calculate the offset $\delta\!\left(\Delta \beta\right)$ on a regular grid in latitude and longitude as
\begin{equation}\label{eq:effective_phase_offset}
    \tan \left[ \delta(\Delta \beta) \right] = \frac{\sum\limits_{-\Delta \beta \leq \beta_j \leq \Delta \beta} \; \sum\limits_{-90^\circ \leq \phi_k \leq 90^\circ} F_{jk} \cos \beta_j \sin \phi_k}{\sum\limits_{-\Delta \beta \leq \beta_j \leq \Delta \beta} \; \sum\limits_{-90^\circ \leq \phi_k \leq 90^\circ} F_{jk} \cos \beta_j \cos \phi_k}
\end{equation}
where $\beta_j$ is the latitude of grid row $j$, $\phi_k$ is the longitude of grid column $k$ relative to the sub-stellar longitude, and $F_{jk}$ is the emission in that cell. The inner sum runs over the day side, and the outer sum over the rows within $\Delta \beta$ in latitude of the equator. The factor $\cos \beta_j$ weights each cell by its area on the sphere. The inset within the eclipse in Figure \ref{fig:GCM_curves_and_maps} shows $\delta(\Delta \beta)$ for the GCM map. We see a clear offset that falls off quickly from a maximum of approximately $4^\circ$ in longitude, with an approximate half-width half-maximum in latitude of $\approx 15^\circ$, comparable to the half-width of the band the WD occults ($\approx 15.6^\circ$). To highlight that this offset is driven by these Rossby wave-like effects in the emission, we make a crude division in the spherical harmonic representation of the GCM emission map into
\begin{itemize}
    \item ``phase-only'' components, which are defined similarly to our phase curve model: $\ell \in \{1, \mathrm{even}\}$, $m=0$; and
    \item ``wave-like'' components, which are defined as all harmonics with $\ell \geq 3$, $m>0$.
\end{itemize}
Recall that we are using the convention in \texttt{jaxoplanet} that places the polar axis at the sub-stellar point. While the choice of definition for the wave-like components is somewhat arbitrary, these harmonics contain almost all the structure that drives the effective phase offset (Figure \ref{fig:GCM_curves_and_maps}, inset within eclipse). They also produce the early--late asymmetry: the eclipse light curve of the phase-only components alone is symmetric to within $\approx 0.1$\% of the eclipse depth, compared with $\approx 6$\% for the full GCM map. As we will show, at roughly three times the SNR implied by our inflated error estimates, such an asymmetry is recovered with the eigenmapping model.

\subsection{Eigenmapping Reproduces the Light Curve Asymmetry from an Equatorial Jet at Sufficient SNR}\label{sec:GCM_test_results:eigenmapping}
For our injection and recovery test, we adopt the top-of-atmosphere near-infrared emission predicted from the best-fit GCM for EPIC 2122 B and scale the fluxes uniformly to reproduce the amplitude of WD 1202 B's emission. We also reproduce the spot and dipole model in our code. We fit and compare results from the spot and dipole model, the phase-only eigenbasis, and the full-phase eigenbasis. Each model is first fit to simulated noise at the error level implied by the maximum-likelihood estimate ($\approx \times 1.4$) of the error inflation factor from the orbital fit in Section \ref{sec:model_methods:orbital_fit}. We then repeat the eigenbasis fits with errors at 2, 3, 5, and $10\times$ the signal-to-noise ratio of the initial fits. Based on the results of these test runs, the $3 \times$SNR case is where latitudinal structure is statistically preferred from our criterion in Section \ref{sec:model_methods:model_selection}. The results we report here use this noise level to prioritize illustrating how the model can recover the signal of a physically realistic structure.

\begin{figure*}
\begin{center}
\includegraphics[width=17cm]{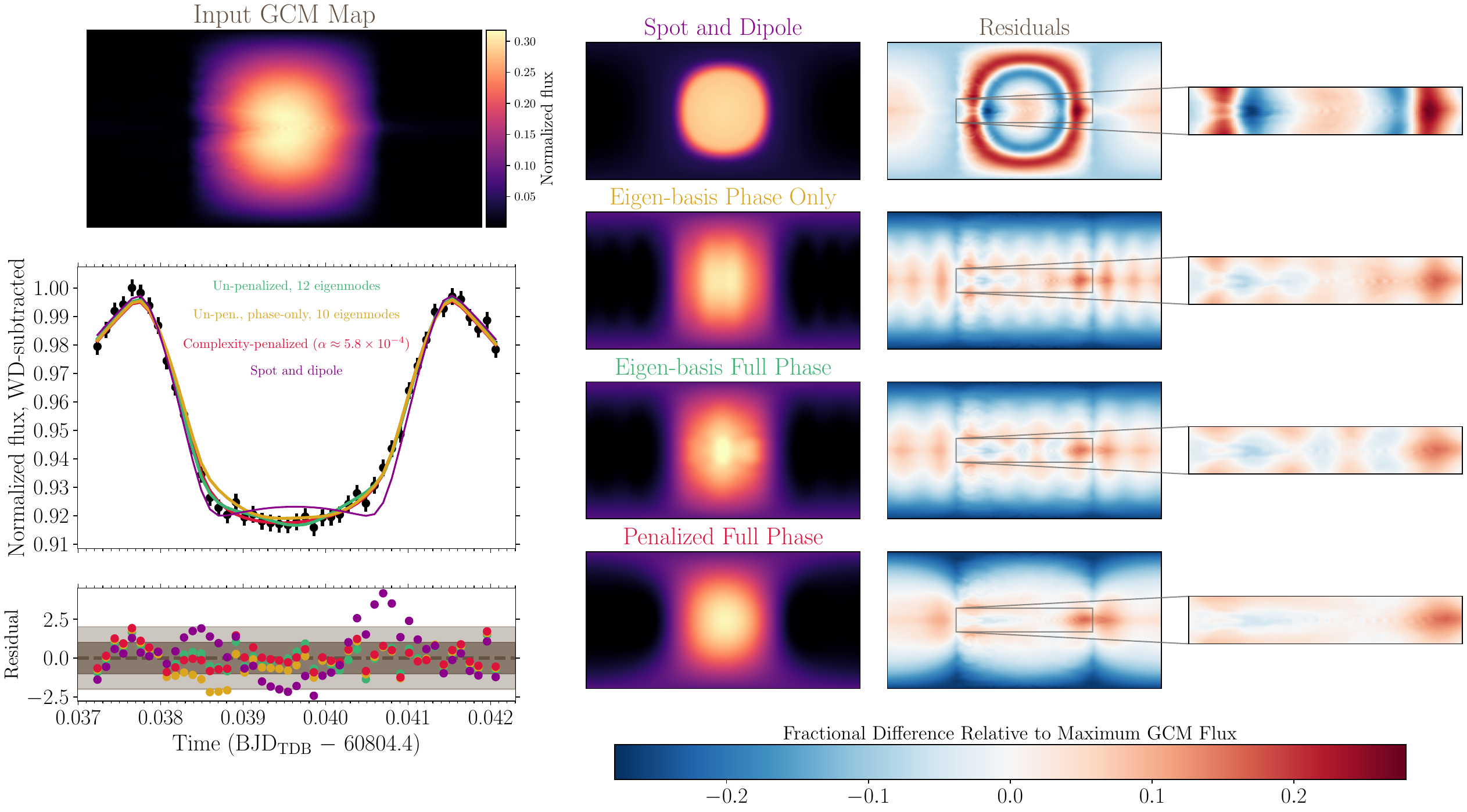}
\caption{A comparison of the best-fit maps from fitting to the EPIC 2122 B-derived GCM map of both (a) the original ``spot and dipole'' model of \citet{Zhou2022a}, originally used to fit the non-eclipsing phase curves of EPIC 2122 B, in purple; and (b) our fit to the eigenbases. For each of these case, we show the fits when only phase data are included (``Phase Only'', in gold), and then with the secondary eclipse included as well (``Full Phase'', in green). The fits to the light curve data in secondary eclipse are shown in the bottom two panels of the left column, while the best-fit emission maps and their residuals (red and blue color scale) are shown as a grid in the remaining columns. The spot-and-dipole model is not able to capture the shape of the eclipse. The fit to the phase-only data with the phase-only eigenbasis improves the fit with a more accurate extrapolated shape in eclipse. However, it under-predicts the eclipse depth in the first half. The fit with the full-phase eigenbasis captures the eclipse shape more accurately by adding structure concentrated along the occultation path. The complexity-penalized model reduces the sharp oscillations in the best-fit emission map.}
\label{fig:GCM_map_comparison_of_fit_methods}
\end{center}
\end{figure*}

Figure \ref{fig:GCM_map_comparison_of_fit_methods} shows the best-fit emission maps, their residuals, and the fits to the light curve in eclipse for the $3 \times$SNR case. The best-fit spot and dipole model returns a hotspot with an angular radius $\approx 58^\circ$, whose profile is relatively flat compared with the flux gradient of the true day side. The map residuals show concentric rings that alternately over- and under-predict the flux. The resulting light curve, while much closer to the GCM light curve, shows a concave shape versus the convex shape of the eclipse data. We infer that the model does not have sufficient freedom to model both the out-of-eclipse phase variations and the eclipse. The fit to the phase-only eigenbasis, which may only use longitudinal variations, fares much better in eclipse. The remaining residuals in eclipse are primarily in the first half, following ingress, where the data dip lower than the same points reflected across the time of mid-eclipse. The phase-only fit fails to capture this asymmetry, but the fit to the full-phase eigenbasis has the flexibility to do so.  Both eigenbases produce fits with oscillations $\sim 20^\circ$ in longitude along the occultation path. The complexity-penalized solution mollifies the sharpness of the oscillations, but none of the models is sensitive to the eastward flux shifts at the peripheries of the day side.

\begin{figure}
\begin{center}
\begin{tabular}{cc}
\includegraphics[height=6cm]{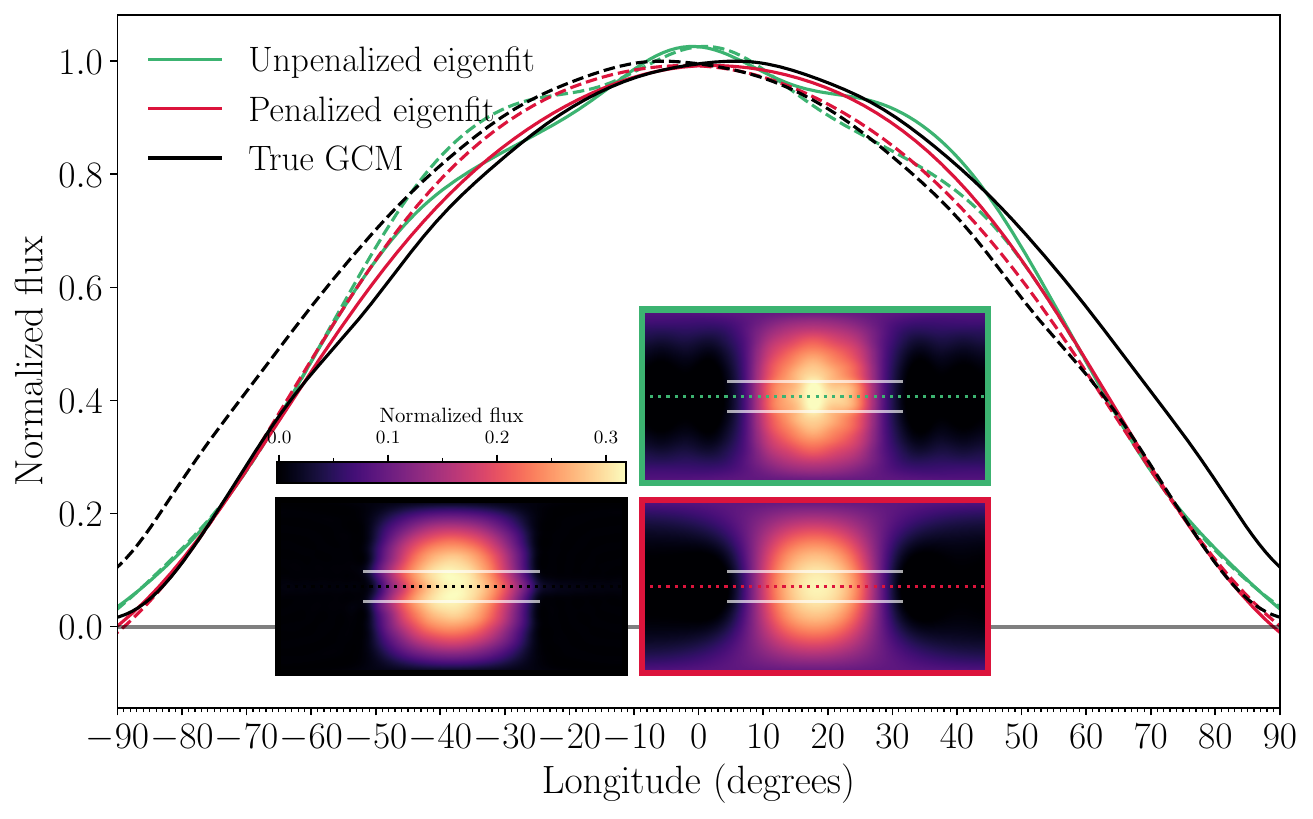} &
\includegraphics[height=6cm]{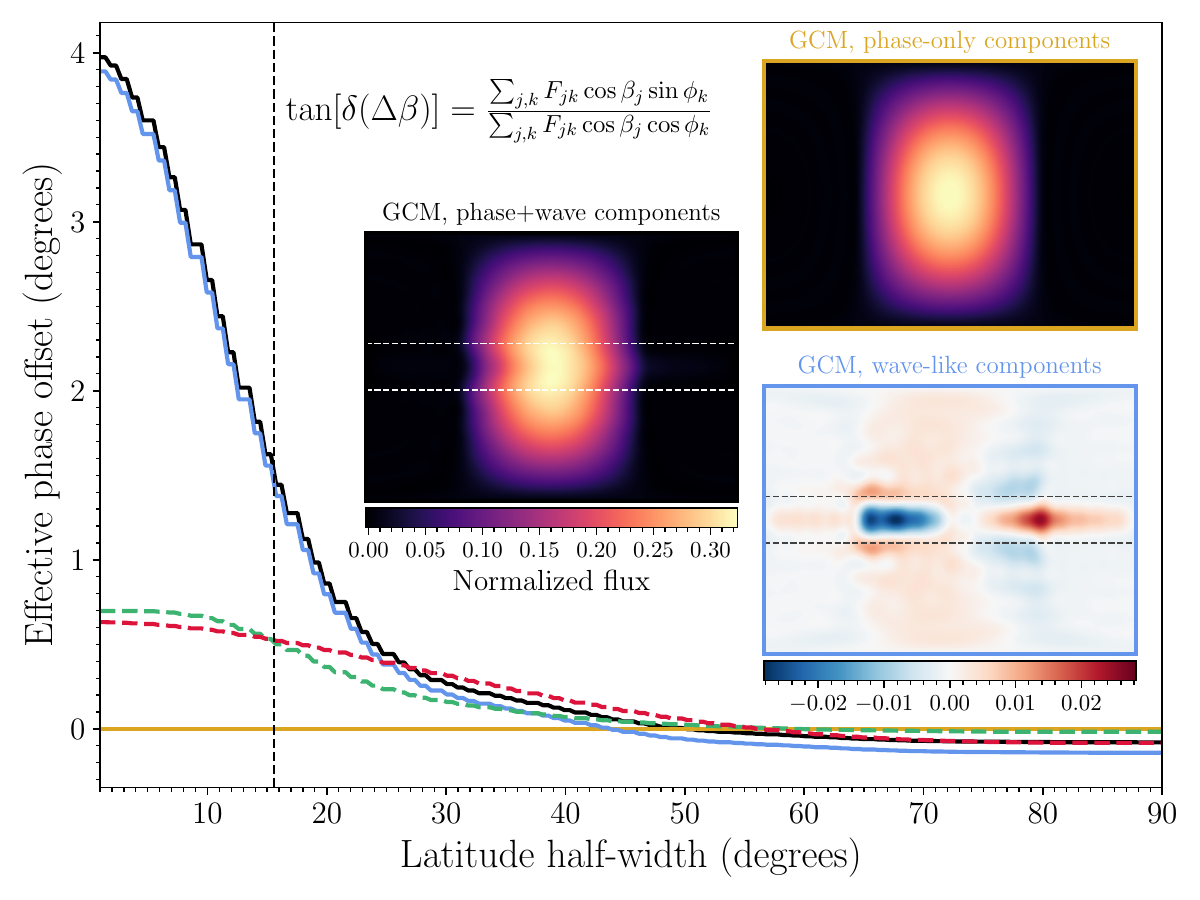} \\
\end{tabular}
\caption{\emph{Left:} Emission fluxes of the best fits to the GCM map introduced in Section \ref{sec:GCM_test_results} at the equator, by longitude. Zero longitude is set to be the sub-stellar longitude. The true fluxes are shown in black, while the fits with the full-phase eigenbasis are shown in green and the complexity-penalized fits (described in Section \ref{sec:model_testing_in_latitude:equatorial_ring_test:complexity_penalization}) in red. \emph{Left, insets:} Maps showing the emission fluxes normalized to the same color scale, with the meridional and zonal slice positions shown as solid and dashed lines, respectively. The parallel light grey lines show the boundaries of where the map is occulted. \emph{Right:} Effective phase offsets ($\delta$) for the model map fits, as well as the true map as a function of the latitude half-width $\Delta \beta$; see Equation \ref{eq:effective_phase_offset}. We also show the phase-only (gold) and wave-like (blue) components of the GCM map defined in Section \ref{sec:GCM_test_results}, as seen in Figure \ref{fig:GCM_curves_and_maps}. The map insets show the components of this decomposition, with the wave-like map plotted on a divergent color scale to show the deviations from the phase-only, day-night contrast components.The dashed vertical line marks the mean half-width of the occulted band ($\approx 15.6^\circ$), and the dashed horizontal lines on the map insets mark the occulted latitudes.}
\label{fig:GCM_penalized_versus_unpenalized_fits}
\end{center}
\end{figure}

The left panel of Figure \ref{fig:GCM_penalized_versus_unpenalized_fits} shows the flux along the equator for both un-penalized and penalized fits, along with the true flux. We see that the penalized model approximates the eastward shift of the flux at the equator. The un-penalized model is similarly biased toward the east, but has a oscillatory pattern on top of the broad latitudinal gradient. The right panel shows the resulting effective phase offset obtained by integrating across successively wider latitude windows, as defined in Equation \ref{eq:effective_phase_offset}. Neither model reproduces the \emph{magnitude} of the true phase offset, but nevertheless successfully recovers a non-zero offset. The penalized profile is broader due to the spreading of the signal to lower-order harmonics. In the true map, by about $50^\circ$, the effective phase offset reaches zero. This implies there is a limit to the latitudes at which the effective phase offset in the retrieved maps stays positive. The spot-like map of the penalized model effectively extends its longitudinally-shifted hotspot as far in latitude as is permitted to reproduce the effective offset.

At this noise level, the statistically preferred fits by BIC use 10 eigenmodes for the phase-only eigenbasis and 12 eigenmodes for the full- eigenbasis. Comparing our model selection criterion (BIC) across the three models, we find the spot and dipole model performs the worst with $\Delta\mathrm{BIC} \approx 241$ relative to the full-phase eigenbasis fit, with the phase-only fit trailing by $\Delta\mathrm{BIC} = 5.8$ (Table \ref{table:GCM_fit_BIC_values}). At this signal-to-noise ratio, both the phase-only and full-phase eigenbases fits are preferred over a spot and dipole fit. The model preference for latitudinal structure relies heavily on the assumed SNR. When the same models are applied to the same GCM-generated light curves, but with the inflated errors from the orbital fit of WD 1202, there is insufficient evidence to prefer the full-phase fit over the phase-only fit. The spot and dipole model is only dis-favored at $\Delta\mathrm{BIC} = 5.2$ relative to the phase-only fit, which is now the most statistically preferred model. Demonstrating eclipse-resolvable structure on this GCM map requires roughly $3\times$ the precision that the WD 1202 data provide. Based on this test alone, if the GCM-predicted jet structure was the primary driver of asymmetry in the eclipse light curve, we would predict that our fits to the real JWST data of WD 1202 may not have sufficient evidence to recover the signal.

\begin{deluxetable*}{rcccc}\label{table:GCM_fit_BIC_values}
\tabletypesize{\scriptsize}
\tablecaption{Model Selection Comparison among Fits to Test GCM Light Curves}
\tablehead{
\colhead{Model Name} & \colhead{Number of Parameters} & \colhead{Mean Brightness ($Y_{0,0}$)} & \colhead{Day-Night Dipole Term ($Y_{1,0}$)} & \colhead{$\mathrm{BIC}-\mathrm{BIC}_\mathrm{min}$}
}
\startdata
Input GCM Map           &       &   0.2615          &   0.2940          &           \\
\cutinhead{\textbf{Inflated Errors ($\times 1.40$)}}
Spot and Dipole         &   4   &   $0.2613\pm0.0003$ & $0.2939\pm0.0003$ &   5.2     \\
Phase-Only Eigenbasis   &   7   &   $0.2818\pm0.0002$ & $0.2935\pm0.0002$ &   0.0     \\
Full-Phase Eigenbasis   &   6   &   $0.2831\pm0.0002$ & $0.2931\pm0.0002$ &   0.7     \\
Penalized               &   N/A &   $0.2832\pm0.0017$ & $0.2937\pm0.0002$ &   N/A     \\
\cutinhead{\textbf{$3\times$SNR}}
Spot and Dipole         &   4   &   $0.2614\pm0.0001$ & $0.2939\pm0.0001$ &   241.0   \\
Phase-Only Eigenbasis   &  11   &   $0.2818\pm0.0001$ & $0.2937\pm0.0001$ &   5.8     \\
Full-Phase Eigenbasis   &  13   &   $0.2812\pm0.0002$ & $0.2938\pm0.0001$ &   0.0     \\
Penalized               &   N/A &   $0.2813\pm0.0020$ & $0.2939\pm0.0001$ &   N/A     \\
\enddata
\end{deluxetable*}

\section{Fits to the Near-Infrared Light Curve of WD 1202}
\label{sec:white_light_fit}
\begin{figure}
\begin{center}
\begin{tabular}{cc}
\includegraphics[height=5.5cm]{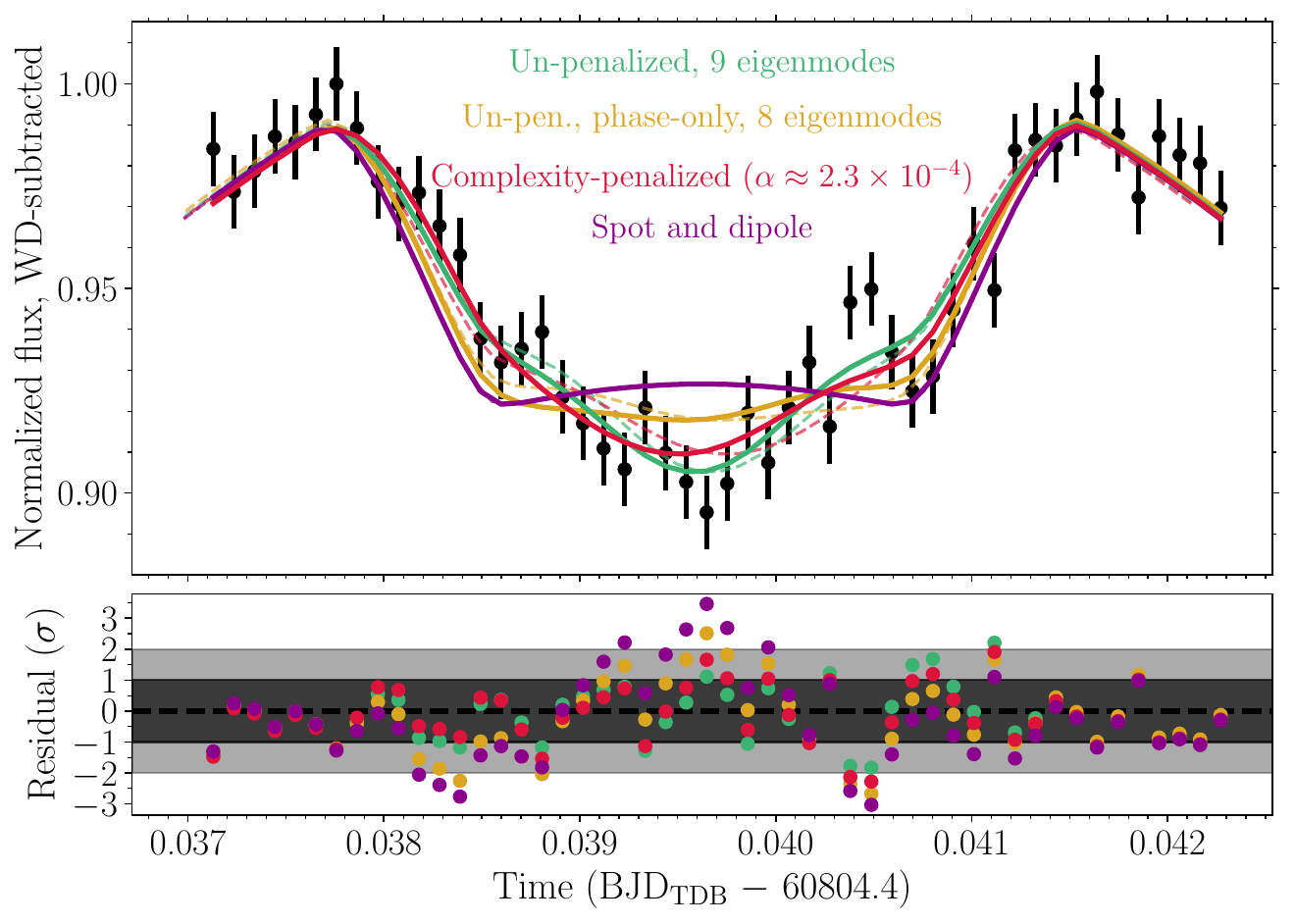} &
\includegraphics[height=5.5cm]{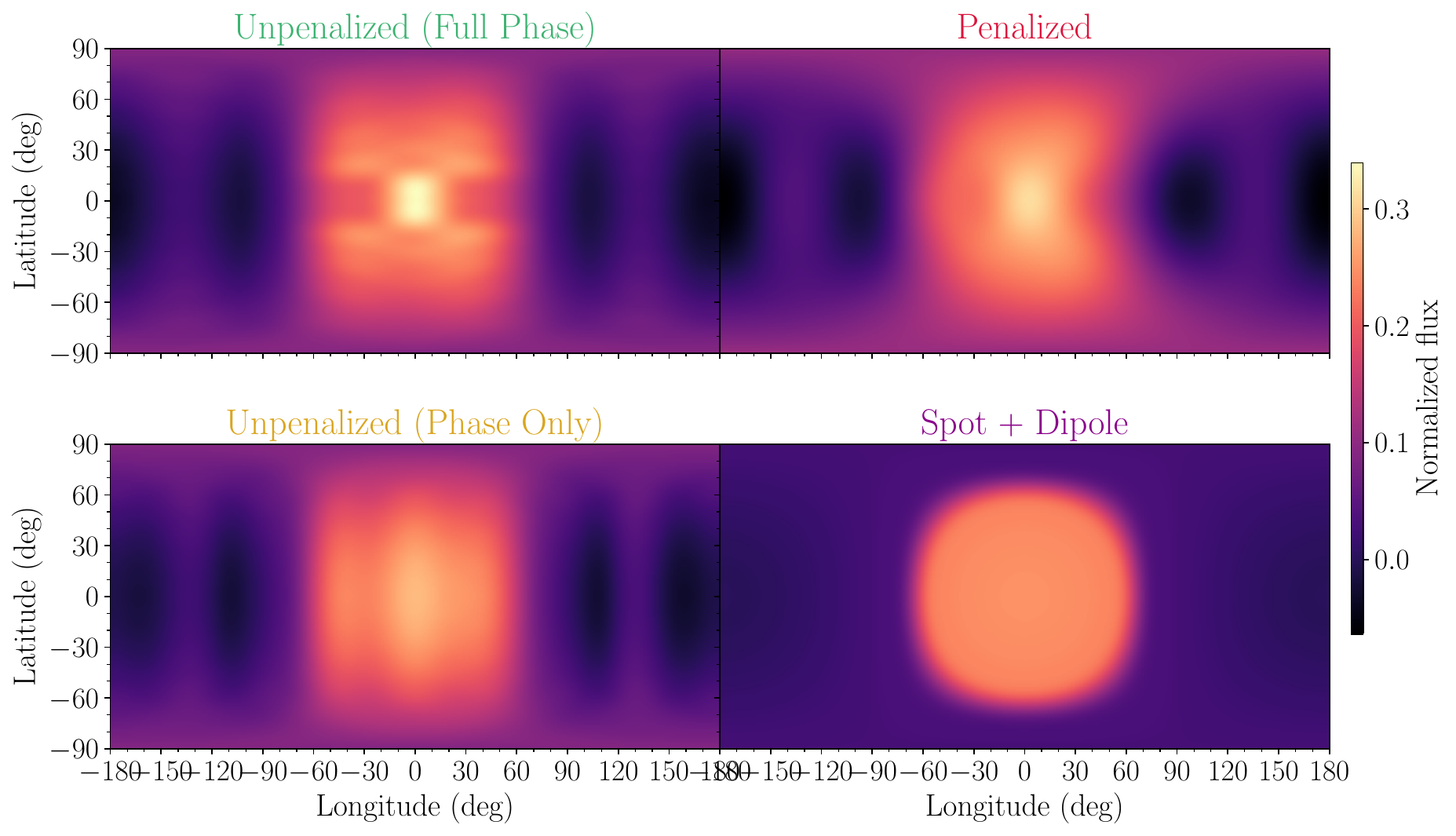} \\
\end{tabular}
\caption{A comparison of the best fit to the wavelength-integrated JWST NIRSpec/PRISM light curve of WD 1202 B, with a model using the eigenbasis (green), versus a fit with a model where the harmonic power is penalized as a function of degree (red, as detailed in Section \ref{sec:model_methods:complexity_penalization}). We select the complexity penalty such that the penalized model's MSE matches that of the eigenbasis fit. \emph{Left:} the models versus data within secondary eclipse, with scaled residuals shown in the lower panel. Here the eclipse is normalized by subtracting the contribution from the WD (as modeled in the initial orbital fit), and dividing by the maximum of the data. \emph{Right}: the best-fit emission maps for the fits to each eigenbasis, the penalized model selected at the $\alpha$ matching the eigenbasis MSE, and a re-fit of the spot and dipole model used in \citet{Zhou2022a}.}
\label{fig:WD1202_white_light_fit}
\end{center}
\end{figure}

\begin{deluxetable*}{rcccc}\label{table:WD1202_fit_BIC_values}
\tabletypesize{\scriptsize}
\tablecaption{Model Selection Comparison among Fits to WD1202 Light Curve}
\tablehead{
\colhead{Model Name} & \colhead{Number of Parameters} & \colhead{Mean Brightness ($Y_{0,0}$)} & \colhead{Day-Night Dipole Term ($Y_{1,0}$)} & \colhead{$\mathrm{BIC}-\mathrm{BIC}_\mathrm{min}$}
}
\startdata
\cutinhead{\textbf{Inflated Errors ($\times 1.40$)}}
Spot and Dipole          &    4 & $0.2616\pm0.0003$      & $0.2651\pm0.0003$      & 62.5     \\
Phase-Only Eigenbasis    &    9 & $0.2788\pm0.0003$      & $0.2651\pm0.0003$      & 27.2     \\
Full-Phase Eigenbasis    &   10 & $0.2800\pm0.0004$      & $0.2650\pm0.0003$      & 0.0      \\
Penalized                &  N/A & $0.2861\pm0.0128$      & $0.2650\pm0.0003$      & N/A      \\
\enddata
\end{deluxetable*}

\begin{figure}
\begin{center}
\begin{tabular}{cc}
  \includegraphics[height=6cm]{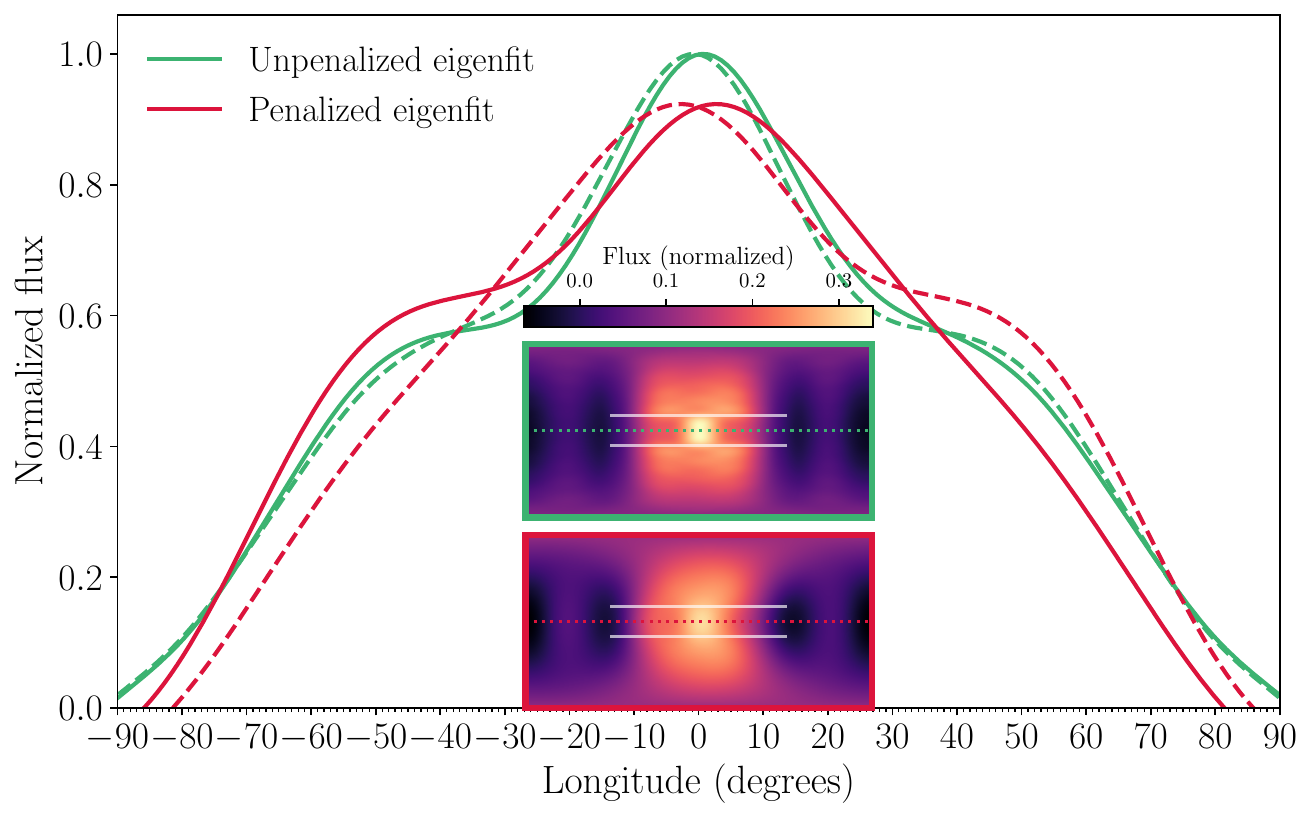} &
\includegraphics[height=6cm]{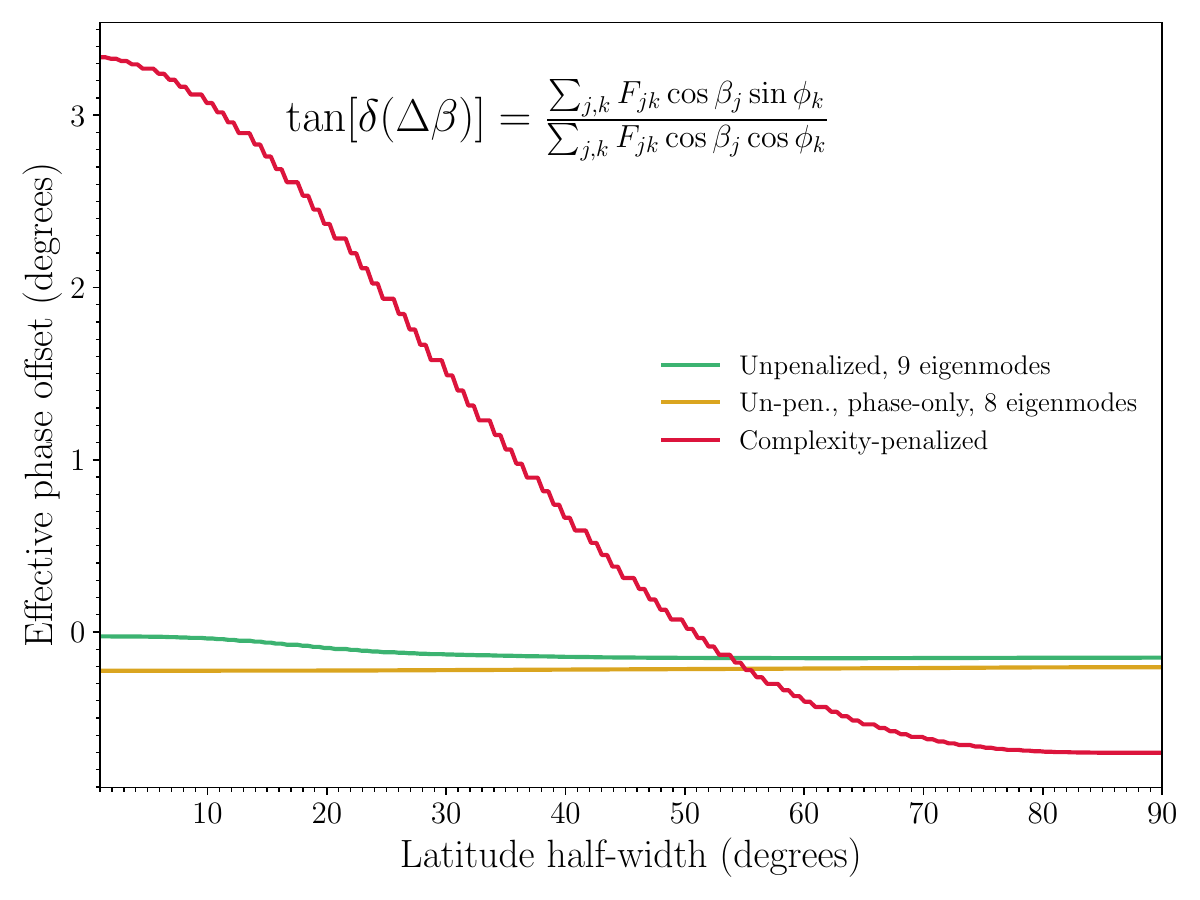} \\
\end{tabular}
\caption{\emph{Left:} Emission fluxes of the best-fit WD 1202 map at the equator, by longitude. Zero longitude is set to be the sub-stellar longitude. The fits with the full-phase eigenbasis are shown in green while the complexity-penalized fits (described in Section \ref{sec:model_testing_in_latitude:equatorial_ring_test:complexity_penalization}) are in red, with dashed curves showing the same slices mirrored about the sub-stellar longitude to highlight any asymmetry. \emph{Insets:} Maps showing the emission fluxes normalized to the same color scale, with the zonal slice position shown as a dotted line. The parallel solid lines show the boundaries of where the map is occulted. \emph{Right:} Effective phase offsets ($\delta$) for the model map fits (as defined in Equation \ref{eq:effective_phase_offset} as a function of the latitude half-width $\Delta \beta$).}
\label{fig:WD1202_penalized_versus_unpenalized_fits}
\end{center}
\end{figure}

We use our models, along with the insights of our injection-recovery tests, to fit and analyze the results of the full-orbit light curve from the JWST NIRSpec/PRISM data of WD 1202. Given the results of our orbital fits in Section \ref{sec:model_methods:orbital_fit}, we limit ourselves to showing only the results for fits to the data with inflated error bars. Figure \ref{fig:WD1202_white_light_fit} shows the in-eclipse light curve fits and the resulting best-fit maps, with the comparison of model statistics, mean brightnesses, and dipole coefficients in Table \ref{table:WD1202_fit_BIC_values}. The full-phase eigenbasis is preferred over the phase-only eigenbasis by $\Delta$BIC $= 27.2$, and over the spot and dipole model by $\Delta$BIC $= 62.5$. The spot and dipole model finds a wide hotspot with an angular radius $\approx 66^\circ$, of similar width to that fit to the GCM map, and recovers an accurate global mean brightness. However, it under-fits the data compared with our other models. Even with inflated errors, there are a few outliers in the data that cannot be reproduced with any model. The two data points closest to the time of mid-eclipse fall below the models by $\sim$ an error bar, as do a pair of data points roughly 2/3 into the eclipse. There are also two points at the beginning of ingress and end of egress that lie at least 2$\sigma$ away from all models.

As with the test GCM fits, the retrieved mean brightnesses are higher in our models than that implied by the initial orbital fit ($\approx 8$--$9\%$ above the phase-curve result of $\approx 0.26$ in normalized flux units). Since the spot-and-dipole result is consistent with that orbital-fit value, our mean brightnesses are likely to be similar to their values rather than those of the eigenmapping fits. The overall shape of the retrieved light curves differs from those of the GCM simulations: our best-fit model maps (Figure \ref{fig:WD1202_white_light_fit}) produce a localized hotspot very close in longitude to the sub-stellar point. Its angular extent is roughly equal in longitude and latitude, at $\approx 15^\circ$. This is due to secondary peaks in the light curve, occurring very roughly about 1/3 and 2/3 of the time across eclipse, that do not resemble the early-late asymmetry profile of the GCM map. The latitudinal extent is confounded with the width (``beam size'') of the projected latitudinal extent of the WD. Figure \ref{fig:WD1202_penalized_versus_unpenalized_fits} compares the equatorial flux profiles of the un-penalized and penalized maps and quantifies their longitudinal asymmetries via the effective phase offset defined in Section \ref{sec:GCM_test_results:effective_phase_offset}: the un-penalized eigenfit places its structure within the occulted band and returns a nearly zero offset at all latitude widths, while the penalized map is broader in latitude and its offset is strongly latitude-dependent, reaching $\approx +3^\circ$ for equatorial bands. We do not have an immediate physical interpretation for the sub-stellar flux concentration in the best-fit maps. While there is a possibility for, say, a variation in the hotspot concentration in angular radius with wavelength, something we would not resolve with our fits to the wavelength-integrated data, a more detailed interpretation is needed, including work to quantify the potential effects of non-astrophysical systematics in the data.

\section{Discussion}
\label{sec:discussion}

\subsection{Regularized Maps in the Context of Hot Jupiters and Isolated Brown Dwarfs}\label{sec:discussion:map_comparison}
When comparing maps across methods and objects, we would like to delineate what the data constrain from what the model retrieves. The eclipse shape, and with it the evidence for latitudinal non-uniformity within the occulted band ($|\beta| \lesssim 16^\circ$), is constrained directly by the light curve. The structure outside that band, the global mean brightness (which we discuss in Section \ref{sec:discussion:mean_brightness_inaccuracy}), and the sharpness of features at the occultation boundary instead depend on the choice of basis and regularization. Regularization makes one choice by filling in the unconstrained structure with the smoothest map consistent with the data at the quality of the fit with the eigenbasis. Maps from other methods are therefore best compared through the quantities the light curve constrains, such as the effective phase offset within the occulted band (Section \ref{sec:GCM_test_results:effective_phase_offset}), rather than the precise structure of the maps. In this sense, our complexity-penalized (``regularized'') solution offers a direct analogy to the longitudinally-offset hotspot maps retrieved from hot Jupiter data in eclipse: as the occulted band widens toward the full day-side hemisphere, the effective phase offset approaches the disk-integrated offset that a hot Jupiter phase curve would measure.

Tikhonov regularization has also recently been employed to map isolated brown dwarfs in rotation: \citet{Wang2026} use regularization to identify spectrally-distinct regions on the brown dwarfs Luhman 16B and SIMP 0136. We do not consider spectral variation in the map structures here, but it is a potentially fruitful next step, particularly since our initial fits to WD 1202's light curve do not follow the predictions of our GCMs. Spectrally-resolved maps may lend insight into the physical mechanisms driving that emission in the infrared. The framework of \citet{Wang2026} could be merged with our approach to perform such an analysis on spectrally-resolved light curves of BDs in short orbits around WDs.

\subsection{Inaccuracies in Retrieved Mean Brightness}\label{sec:discussion:mean_brightness_inaccuracy}
As seen in the fits to the emission from a BD companion simulated from a GCM (Section \ref{sec:GCM_test_results}), while all models return a day-night dipole coefficient correct to within at most twice the estimated model uncertainty, the fits with the eigenbases over-predict the global mean brightness by $\approx 8$\%. The spot and dipole models do not suffer this inaccuracy. The difference in results comes down to how each model concentrates its predicted flux toward the sub-stellar point. Both the initial low-order spherical harmonic model of the non-eclipse orbital fit (Section \ref{sec:model_methods:orbital_fit}) and the spot and dipole model only contain components that are radially symmetric about the sub-stellar point. Therefore, the structures they can use to fit the light curve data preferentially concentrate flux at the sub-stellar point, which is the most efficient way to increase the day-side maximum in the light curve with the least additional emission flux. The eigenbases fits are not restricted in this way, instead localizing the signal strength where we are most sensitive. At high latitudes (as seen in the fit maps of Figure \ref{fig:GCM_curves_and_maps}), the eigenbasis map fits tend to return to the mean flux rather than going to zero, increasing the global mean brightness.

The complexity-penalized solutions also have global mean brightnesses inconsistent with the true value, but return much larger uncertainties than the eigenbases fits. Though not recorded in Table \ref{table:GCM_fit_BIC_values}, there is a nearly perfect anti-correlation between the constant mode/harmonic with each of the sub-stellar zonal quadrupole (in \texttt{jaxoplanet}, $Y_{2,0}$) and the longitudinal sectoral quadrupole ($Y_{2,2}$). $Y_{2,0}$ and $Y_{2,2}$ are nearly perfectly positively correlated with each other, and --- not coincidentally --- both are important in the spherical harmonic decomposition of the equatorial ring (Section \ref{sec:model_testing_in_latitude:equatorial_ring_test}). In effect, then, there is a degeneracy between the hemispheric brightness and the amount of equatorial concentration. It is the sectoral harmonic $Y_{2,2}$ that requires a higher mean brightness to add to the day-side light curve signal than either of $Y_{2,0}$ and $Y_{0,0}$, and is the principal source of the inaccuracy.

One may then ask, if simpler models like the spot and dipole can retrieve the mean brightness accurately, whether one can then fix the mean brightness in the eigenmapping models to the spot and dipole result. We test this on the simulated GCM cases without error inflation and find that this degrades the quality of the fit such that the BIC is comparable with that of the spot and dipole model. This is because of the structure of the eigenmodes: by construction, all eigenmodes beyond the constant mode impart zero change to the mean brightness, and in this case, if the mean brightness is fixed to a level that is not optimized for the non-constant eigenmodes, we cannot tune the coefficients to produce a fit matching the quality of the model with a free constant term.

We also note that this normalization-level bias is distinct from the under-recovery of the occultation asymmetry metric $g$ in our injection tests. $g$ is defined in a way that is insensitive to the global mean brightness, and its failure to converge to the true value in both the dipole and ring cases therefore reflects the degeneracy of the maps that our model finds in fitting the light curves, as discussed in Section \ref{sec:model_testing_in_latitude:equatorial_ring_test:eigenmapping}.

\subsection{Uncertainties in Impact Parameter}\label{sec:discussion:impact_parameter_uncertainty}
For this work we have assumed the maximum-likelihood value ($b \approx 0.015$) for both our simulations and fits to the WD 1202 data. Recall from the initial orbital fit (Table \ref{table:orbital_fit_results}) that the 68\% credible interval reaches as high as $b \approx 0.98$, with little correlation between the impact parameter and other parameters (see Appendix \ref{sec:appendix:orbital_fit_corner_plot} for the full corner plot of each parameter). We highlight the upper bound of the interval since this weakens the potential constraint on signals from the equatorial region, where the effects of super-rotation impart the time asymmetry in the light curve in eclipse. A lack of recovery of an early-late asymmetry in our data, were it to exist, is likely to be primarily caused by insufficient SNR. However, another factor could be a significantly non-zero impact parameter --- one that weakens the imparted signal from the jet in the eclipse data. As we conclude in Section \ref{sec:white_light_fit}, the interpretation is still unclear.

\section{Conclusions}
\label{sec:conclusions}
This work presents an initial study of the constraints on the spatial brightness structure of sub-stellar companions around white dwarf hosts when the systems are eclipsing, where only partial occultations of the companion can be observed. We adapt the eigenmapping technique, rooted in principal component analysis, to construct a basis of emission maps whose light curves impart the most signal across the companion's rotation and secondary eclipse. We then augment the model to achieve our final map fit by imposing a penalty on the underlying harmonic structure of the fits via Tikhonov regularization. This represents one method to define a ``simplest'' solution to the emission from the light curve. We use the eclipsing BD--WD system WD 1202 --- with full-phase observations from JWST NIRSpec --- as a template for a set of sensitivity tests of partial eclipse mapping. As a preparatory step for eigenmapping, we use an initial fit with low-order spherical harmonics to all light curve data except those in secondary eclipse to constrain the orbital geometry, and find that our model estimates the data uncertainties to be under-predicted on average by a factor of $\approx 1.4$.

As a precursor to our fit to the JWST data of WD 1202, we construct and run a series of injection and recovery tests. These start with very simple, artificial latitudinal structures --- a north-south dipole map, and a map with emission concentrated in a ring at the equator --- followed by a more realistic artificial data set generated from the near-infrared emission of a GCM simulation of a similar BD--WD system EPIC 2122. We then apply the model to the wavelength-integrated full-orbit JWST NIRSpec phase curve of WD 1202. We find
\begin{itemize}
    \item While eigenmapping can effectively detect latitudinal non-uniformities within the occultation path, our tests reveal that the fits over-estimate the global mean brightness for maps with flux concentrated toward the equator. This bias occurs because eigenmapping localizes structure within the occultation path and otherwise defaults to a constant brightness. As a result, the model cannot distinguish between a genuinely brighter uniform map and the projection effect of an equatorial band, leading it to overestimate the flux in unconstrained regions like at high latitudes.
    \item Tikhonov regularization provides a smoothly varying global map whose fit quality can be directly tuned to the statistically-optimized eigenmapping fit; we use this model as our preferred estimate of the emission map of the BD. While this complexity penalization can successfully recover low-degree harmonic structures like a north-south dipole, its suppression of high-degree harmonics prevents it from accurately resolving sharp, harmonically complex latitudinal boundaries such as those of an equatorial ring. This is by design, and provides a representation of our knowledge of the spatial scales of structures in emission.
    \item We show that our models can detect and recover the time asymmetry produced by narrow equatorial Rossby waves in a GCM simulation of a highly-irradiated, rapidly-rotating BD, but only for simulations that have artificial white noise at $\sim 3\times$ the signal-to-noise ratio of those we estimate for the JWST data for WD 1202 from our orbital fit. Our regularized model represents the asymmetry as a latitude-dependent ``effective phase offset'' that provides an analog to phase offsets in hot Jupiter light curves.
    \item Our model of the wavelength-integrated JWST NIRSpec/PRISM orbital light curve of WD 1202 yields a fit that statistically prefers latitudinal non-uniformity in the map. However, unlike the predictions from the GCM simulations, our recovered structure shows a secondary concentration of flux on the day side centered at the sub-stellar point with angular scale $\approx 15^\circ$. This empirical result does not yield an immediate physical interpretation and therefore motivates a follow-up analysis.
\end{itemize}

This Tikhonov-regularized eigenmapping approach represents just one possible reconstruction of a global map for partially occulting systems. That being said, it now has a nearly-parallel application in sub-stellar rotational mapping, as recently demonstrated in \citet{Wang2026}. It relies on tuning the spatial penalty based on the discrete statistical model optimization of the eigenmapping model, so that we can compare their map structures. Other complementary eclipse mapping frameworks employ more direct optimization techniques, for example the cross-validation methods developed in \citet{Hammond2024} for hypothesis testing and complexity tuning in pixel space. Applying cross-validation to partial-occultation mapping in future work could provide an absolute optimization of the complexity penalty. Additionally, with further adaptations, the techniques developed here for partial-occultation mapping could have broader implications for exoplanet science beyond just BD--WD binaries. For instance, a similar mapping framework could theoretically be used to map surface heterogeneities in host stars during transits in multi-planet systems, potentially providing an empirical constraint on the transit light source effect \citep[e.g.][]{Rackham2018}.

The code used in this work is available in a public repository under the name \href{https://github.com/adadams/iluna}{\texttt{ILUNA}}. The repository is currently in preparation for release as a publicly-available package for reproducing the analyses including and beyond those presented in this work.

\appendix
\section{Posterior Distributions for the Keplerian Orbital Fit of WD 1202}
\label{sec:appendix:orbital_fit_corner_plot}
\begin{figure}[htb!]
\begin{center}
\includegraphics[width=17cm]{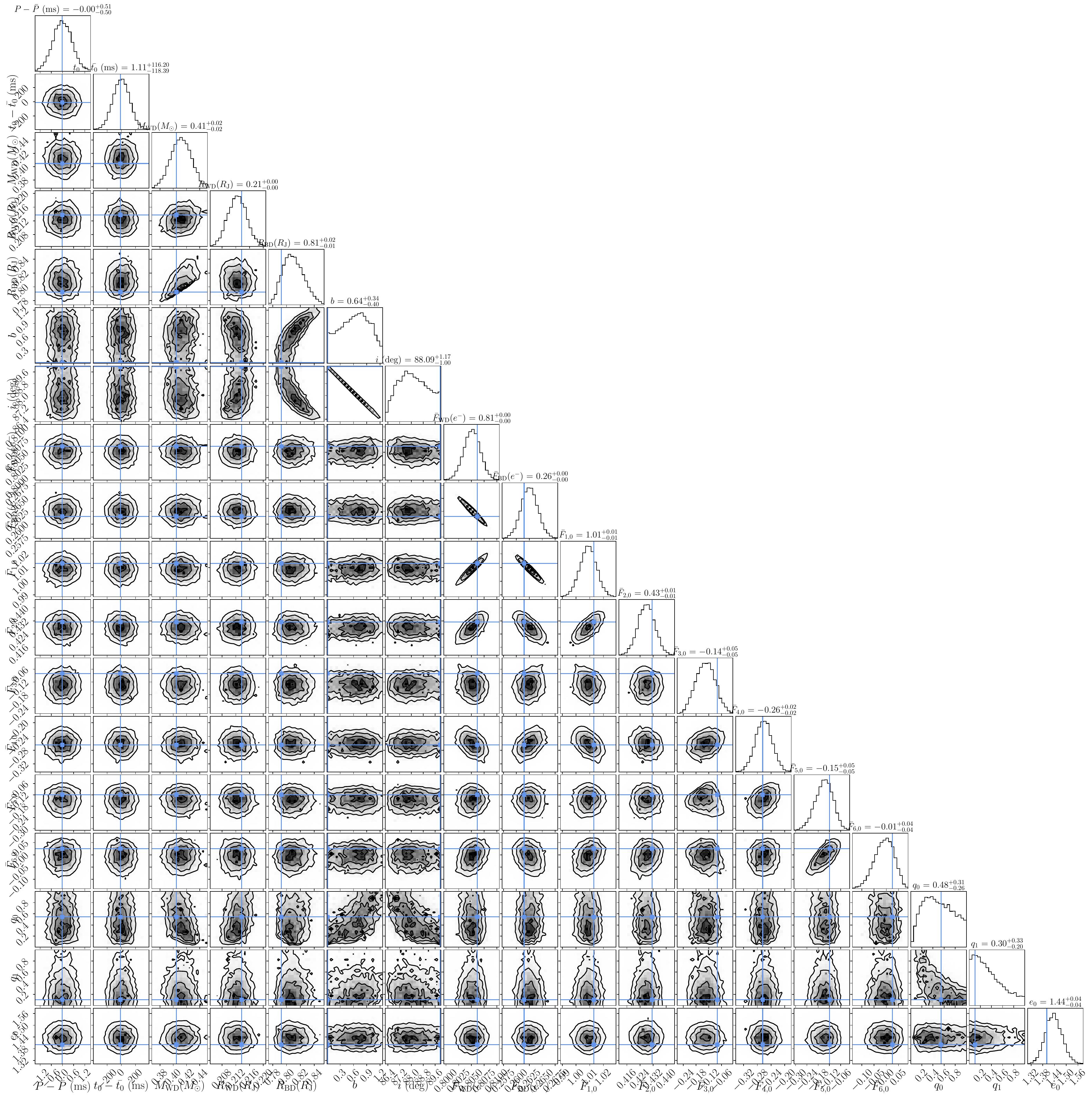}
\caption{A corner plot of the posterior distributions of the orbital fit to the wavelength-integrated JWST NIRSpec/PRISM light curve of WD 1202, excluding data in secondary eclipse. The priors, maximum likelihood estimates, and credible intervals of the free parameters are listed in Table \ref{table:orbital_fit_results}. Here the orbital inclination $i$ is not a free parameter and is instead calculated a posteriori from the impact parameter and other orbital elements.}
\label{fig:WD1202_orbital_fit_corner_plot}
\end{center}
\end{figure}

\begin{acknowledgements}
This work is based on observations made with the NASA/ESA/CSA James Webb Space Telescope. The data were obtained from the Mikulski Archive for Space Telescopes at the Space Telescope Science Institute, which is operated by the Association of Universities for Research in Astronomy, Inc., under NASA contract NAS 5-03127 for JWST. These observations are associated with program \#04967. Support for program \#04967 was provided by NASA through a grant from the Space Telescope Science Institute, which is operated by the Association of Universities for Research in Astronomy, Inc., under NASA contract NAS 5-03127. YZ acknowledges support from a Heising-Simons Foundation 51 Pegasi b Alumni Faculty Grant (\#2023-4808).

We used the following AI tools in preparing this work: Cognition SWE-2 and Anthropic Claude Opus 5.5. The design of the \texttt{iluna} code framework, its initial implementation, and the text of this manuscript are the authors' own. AI tools were used to write and revise portions of the analysis and plotting scripts; to bring that code into a consistent style with the authors' previous code; to check consistency among the code, figures, and text; and to suggest wording for some passages which the authors rewrote. All AI-assisted code was reviewed by the authors and validated through the injection-recovery tests of Sections \ref{sec:model_testing_in_latitude} and \ref{sec:GCM_test_results}, which recover known input maps; and by regenerating every figure from the public code when scripts were revised, and confirming that the outputs did not change.

\end{acknowledgements}

\facilities{JWST (NIRSpec)}

\bibliography{library}{}
\bibliographystyle{aasjournalv7}



\end{document}